\documentclass[12pt]{iopart}

\begin{document}

\title{Equations of a Moving Mirror and the Electromagnetic Field}

\author{Luis Octavio Casta\~{n}os$^{1}$ and Ricardo Weder$^{2}$}

\address{Departamento de F\'{i}sica Matem\'{a}tica, Instituto de Investigaciones en Matem\'{a}ticas Aplicadas y en Sistemas, Universidad Nacional Aut\'{o}noma de M\'{e}xico, Apartado Postal 20-126, M\'{e}xico DF 01000, M\'{e}xico}

\ead{1. loccj@yahoo.com \ , 2. weder@unam.mx}
\vspace{10pt}
\begin{indented}
\item[]
\end{indented}

\begin{abstract}
We consider a system composed of a mobile slab and the electromagnetic field. We assume that the slab is made of a material that has the following properties when it is at rest: it is linear, isotropic, non-magnetizable, and ohmic with zero free charge density. Using instantaneous Lorentz transformations, we deduce the set of self-consistent equations governing the dynamics of the system and we obtain approximate equations to first order in the velocity and the acceleration of the slab. As a consequence of the motion of the slab, the field must satisfy a wave equation with damping and slowly varying coefficients plus terms that are small when the time-scale of the evolution of the mirror is much smaller than that of the field. Also, the motion of the slab and its interaction with the field introduce two effects in the slab's equation of motion. The first one is a position- and time-dependent mass related to the \textit{effective mass} taken in phenomenological treatments of this type of systems. The second one is a velocity-dependent force that can give rise to friction and that is related to the much sought \textit{cooling} of mechanical objects.
\end{abstract}

%
%
\submitto{\PS}
%
\maketitle
%
%

\section{INTRODUCTION} 

Optomechanics studies systems composed of \textit{light} (the electromagnetic field) and mechanical objects such as movable membranes or mirrors \cite{Trends}-\cite{Nuevo2}. The interaction between the two components is governed by two principal forces: radiation pressure and thermal forces. The latter are also called \textit{bolometric forces} and consist in light absorption deflecting the mechanical object \cite{Restrepo}. In general, both forces are always present, but, depending on the optomechanical system, one can be much larger than the other and they can point in different directions \cite{Restrepo}. Using certain experimental setups and certain types of materials one can manage to have mechanical objects in which one force predominates over the other. In this way, one has systems in which radiation pressure dominates \cite{Otro,Kip1, Kip2}, others in which thermal forces are much larger \cite{Mar1, Mar2}, and some others in which both types of forces are comparable \cite{Italia1}-\cite{Israel2}.  

Research in optomechanics is driven mainly by technological applications in areas such as lightwave communications \cite{Israel0} and studies in classical and quantum physics. In particular, optomechanical systems exhibit intricate classical non-linear dynamics \cite{Kip1}-\cite{Israel2}, \cite{NuestroPRA,NuestroIOP} and may be a setting where quantum physics can be studied in the macroscopic domain \cite{Trends}-\cite{Nuevo2}. A first step that may open the door for such research is that mechanical objects have been cooled to their ground state in some experimental set ups \cite{Nature1, Nature2}.

In this work we study a system composed of the electromagnetic field and a mobile mirror in the form of a slab. Using the transformation properties of electromagnetic quantities under Lorentz transformations and not considering thermal effects, we establish the classical equations that determine the evolution of the system. Afterwards, we deduce equations that are correct up to first order in both the velocity and acceleration of the slab. With these results at hand, we conclude that the motion of the mirror and its interaction with the field give rise to a position- and time-dependent mass related to the \textit{effective mass} taken in phenomenological treatments of this type of systems \cite{Nuevo2,Mar2,Israel1} and to a velocity-dependent force that is related to the \textit{cooling} of mechanical objects \cite{Nuevo2,Restrepo}. Moreover, the field must satisfy a wave equation that depends on the slab's position, velocity, and acceleration.

There are a fair number of works discussing the equations governing these type of systems, see \cite{Law1,Law2} and references therein. In particular, \cite{Law1} considers a one-dimensional cavity composed of one perfect, fixed mirror and one perfect, mobile mirror with empty space in between, while \cite{Law2} considers a one-dimensional cavity composed of two perfect, fixed mirrors and one mobile, non-conducting mirror with constant electric susceptibility in between. Both \cite{Law1,Law2} have the objective of deducing a Hamiltonian that approximately describes the mobile-mirror + electromagnetic field system and give a justification of the Hamiltonian usually used in the area of quantum optomechanics. Our work is completely different in spirit, since our objective is to establish the exact classical equations governing the system from first principles and to obtain consistent approximations of them. Furthermore, we consider that the slab is an ohmic conductor and we establish our equations for general electromagnetic fields and dielectric and conductivity functions. Moreover, we deduce the time-dependent mass and velocity-dependent force affecting the motion of the mirror mentioned above, quantities that cannot be deduced correctly from the approximate Lagrangian used in \cite{Law2} because it does not give the correct equation of motion for the mirror to first order in its velocity and acceleration, as it is mentioned in \cite{Law2}.

Finally, we have strived in keeping the treatment simple enough, since only an elementary knowledge of electromagnetism and special relativity is required.

The material is organized as follows: In Sec. II we introduce the system under study and we establish the model used to describe it. In Sec. III we determine the equations governing the evolution of the electromagnetic field, while Sec. IV considers the special case of a linearly polarized electric field and obtains approximations to the equation governing the evolution of the field. In Sec. V we deduce the force affecting the motion of the mobile mirror and establish the equations governing the evolution of the complete field-mobile mirror system. In Sec. VI we discuss the validity of the model and in Sec. VII we present a Lagrangian density for the electromagnetic field. Finally, the conclusions are given in Sec. VIII. Part of this work was presented, without giving any details of the deduction of the results, at the \textit{Latin America Optics and Photonics Conference (LAOP) 2014} held at Canc\'{u}n, M\'{e}xico \cite{LAOP1}.


\section{THE SYSTEM}

The system under study is composed of a mobile mirror and the electromagnetic field. For simplicity, we assume that the mirror has the form of a slab that has infinite length and width and that has thickness $\delta_{0}$ when it is at rest. Moreover, we assume that the mirror is made of a material that has the following properties when it is at rest (we call them the \textit{rest properties of the mirror}): it is linear, isotropic, non-magnetizable, and ohmic with zero free charge density, electric susceptibility $\chi$, and conductivity $\sigma$. In all that follows (unless otherwise stated), we assume that the electric susceptibility and conductivity have continuous first derivatives in all of space and that $\chi$ and $\sigma$ are equal to zero outside of the mirror. Also, we use the Minkowski metric (that is, the $ict$-system) for special relativity and Gaussian units for the electromagnetic field.

We mention that it is important to distinguish which properties are satisfied when the mirror is at rest, since it will be shown explicitly in the following sections that these do not hold when the mirror is in motion.

We want to establish the equations governing the evolution of the system in an inertial reference frame \textbf{LS} (the \textit{Laboratory System}) where
\begin{enumerate}
\item A right-handed Cartesian coordinate system is specified by unit vectors $\mathbf{\hat{x}}$, $\mathbf{\hat{y}}$, and $\mathbf{\hat{z}}$ along the positive directions of the coordinate axes.

\item The coordinates of an arbitrary event are denoted by $(x,y,z,ict)$. Also, $\mathbf{r} \equiv (x,y,z)$ and $c$ is the speed of light in vacuum.

\item The mirror can move only along the $x$-axis and it fills the region
\begin{eqnarray}
\label{1}
R(t) \ = \ \left\{ \ \mathbf{r}\in\mathsf{R}^{3}: \ \left\vert x-q(t) \right\vert \leq \frac{\delta(t)}{2} \right\} .
\end{eqnarray}
Here $q(t)$ is the position of the midplane of the mirror along the $x$-axis, while $\delta (t)$ is the mirror's thickness in the $x$ direction. Observe that $\delta (t)$ is time-dependent because the mirror can be in motion and, consequently, it can be Lorentz contracted. In all that follows we call $q(t)$ the midpoint and $\delta (t)$ the thickness of the mirror in $\textbf{LS}$. Also, $\dot{q}(t) = (dq/dt)(t)$ and $\ddot{q}(t) = (d^{2}q/dt^{2})(t)$ are the velocity and acceleration of the midpoint $q(t)$, respectively.

\item Outside of $R(t)$ there is vacuum.

\item We assume that all electromagnetic quantities are functions of only $x$ and $t$. Moreover, $\mathbf{E}(x,t)$ denotes the electric field, $\mathbf{B}(x,t)$ the magnetic field, $\mathbf{P}(x,t)$ the polarization, $\mathbf{M}(x,t)$ the magnetization, $\mathbf{J}_{f}(x,t)$ the free current density, and $\rho_{f}(x,t)$ the free charge density. 
\end{enumerate}

Now the goal is to deduce the form of $\mathbf{P}(x,t)$, $\mathbf{M}(x,t)$, $\mathbf{J}_{f}(x,t)$, and $\rho_{f}(x,t)$. It is tempting to calculate these quantities by assuming that the rest properties of the mirror are valid even if the mirror is in motion, especially in the case of small velocities. Nevertheless, it is shown explicitly in the following sections that one neglects important phenomena if one decides to take that approach. Therefore, we take another path to determine the aforementioned quantities in \textbf{LS}. We now describe the idea behind our approach. At each instant of time in \textbf{LS} we consider another inertial reference frame \textbf{MS} where the midpoint $q(t)$ of the mirror is instantaneously at rest. If the mirror is not subject to very large accelerations, the mirror will be approximately at rest in \textbf{MS} during a small time interval. Hence, one can use the rest properties of the mirror to calculate the polarization, magnetization, free current density and free charge density in \textbf{MS} during this small time interval. Afterwards, one can use their well known transformation properties under Lorentz transformations to obtain these quantities in \textbf{LS}. A word of caution, this \textit{instantaneous Lorentz transformation} works only if both the coordinate origin and the origin of time in \textbf{LS} are first translated to be at $q(t)$ and at the instant of time one is considering.  We now formalize this idea. In particular, we determine quantitatively the aforementioned \textit{small time interval} and we establish conditions under which the mirror is subject to small accelerations.

Let $t_{0} \in \mathsf{R}$ be fixed. Consider an inertial reference frame \textbf{LS}$_{0}$ obtained from \textbf{LS} by a time and space translation where $t_{0}$ is the new origin of time and $(q(t_{0}),0,0)$ is the new origin of space. In \textbf{LS}$_{0}$ one has the following properties:
\begin{enumerate}
\item A Cartesian coordinate system is specified by unit vectors $\mathbf{\hat{x}}'$, $\mathbf{\hat{y}}'$, and $\mathbf{\hat{z}}'$ parallel respectively to $\mathbf{\hat{x}}$, $\mathbf{\hat{y}}$, and $\mathbf{\hat{z}}$.

\item The coordinates of an arbitrary event are denoted by $(x',y',z',ict')$ and they are related to the corresponding coordinates $(x,y,z,ict)$ in \textbf{LS} by
\begin{eqnarray}
\label{2}
x' &=& x-q(t_{0}) \ , \ \ \ \ y'  = y \ , \cr 
t' &=& t-t_{0} \ , \ \ \ \ \ \ \ \ z' = z \ .
\end{eqnarray}
Also, $\mathbf{r}' \equiv (x',y',z')$.

\item $\mathbf{E}'(x',t')$ denotes the electric field, $\mathbf{B}'(x',t')$ the magnetic field, $\mathbf{P}'(x',t')$ the polarization, $\mathbf{M}'(x',t')$ the magnetization, $\mathbf{J}_{f}'(x',t')$ the free current density, and $\rho_{f}'(x',t')$ the free charge density. Notice that a quantity $f'(x',t')$ in \textbf{LS}$_{0}$ is connected to the corresponding quantity $f(x,t)$ in \textbf{LS} by
\begin{eqnarray}
\label{3p3}
f'(x',t') &=& f\left[ x' + q(t_{0}),t'+t_{0} \right] \ .
\end{eqnarray}
\end{enumerate}
\noindent
Observe that at time $t'$ the mirror fills the region
\begin{eqnarray}
\label{3}
R'(t') &=& \left\{ \ \mathbf{r}'\in\mathsf{R}^{3}: \  \left\vert x'-q'(t') \right\vert \leq  \frac{\delta'(t')}{2} \right\} \ ,
\end{eqnarray}
with
\begin{eqnarray}
\label{midpointLS0}
q'(t') = q(t'+t_{0}) -q(t_{0}) \ , \ \ \ \delta'(t') = \delta(t'+t_{0}) \ , 
\end{eqnarray} 
its midpoint and its thickness along the $x'$-axis. As in the case of $q(t)$ and $\delta (t)$, we call $q'(t')$ the midpoint and $\delta'(t')$ the thickness of the mirror in \textbf{LS}$_{0}$. Also, there is vacuum outside of $R'(t')$. 

In particular, at time $t'=0$ the mirror fills the region
\begin{eqnarray}
\label{4}
R'(0) \ = \ \left\{ \ \mathbf{r}'\in\mathsf{R}^{3}: \ \left\vert x' \right\vert \leq  \frac{\delta(t_{0})}{2} \ \right\} \ ,
\end{eqnarray}
and its mid-point satisfies
\begin{eqnarray}
\label{midpointLS00}
q'(0) = 0 \ , \ \ \frac{dq'}{dt'}(0) = \dot{q}(t_{0}) \ , \ \ \frac{d^{2}q'}{dt'^{2}}(0) = \ddot{q}(t_{0}) \ .
\end{eqnarray} 
Now that we have established a reference frame in which the coordinate origin is at the midpoint of the mirror and the time origin has been redefined appropriately, we introduce another reference frame in which the mirror is instantaneously at rest.

Define
\begin{eqnarray}
\label{5}
v_{0} = \dot{q}(t_{0}) \ , \ \ \ \beta_{0} = \frac{v_{0}}{c} \ , \ \ \ \gamma_{0} = \frac{1}{\sqrt{1-\beta_{0}^{2}}}  \ .
\end{eqnarray}
Notice that $v_{0}$ and, consequently, $\beta_{0}$ can be positive or negative.  

Consider an inertial reference frame \textbf{MS}$_{0}$ (for \textit{Mirror System}) where
\begin{enumerate}
\item A Cartesian coordinate system is specified by unit vectors $\mathbf{\hat{x}}''$, $\mathbf{\hat{y}}''$, and $\mathbf{\hat{z}}''$ parallel respectively to $\mathbf{\hat{x}}'$, $\mathbf{\hat{y}}'$, and $\mathbf{\hat{z}}'$.

\item \textbf{MS}$_{0}$ moves with velocity $v_{0}\mathbf{\hat{x}}'$ with respect to \textbf{LS}$_{0}$.

\item The coordinates of an arbitrary event are denoted by $(x'',y'',z'',ict'')$ and they are related to the corresponding coordinates $(x',y',z',ict')$ in \textbf{LS}$_{0}$ by a Lorentz transformation:
\begin{eqnarray}
\label{7}
ct'' &=& \gamma_{0}(ct' - \beta_{0}x') \ , \qquad y'' \ = \ y' \ , \cr
x'' &=& \gamma_{0}(x' - \beta_{0}ct') \ ,  \qquad z'' \ = \ z' \ .
\end{eqnarray}
Here the space-time origin $(\mathbf{r}'=\mathbf{0},ict'=0)$ of \textbf{LS}$_{0}$ coincides with the space-time origin $(\mathbf{r}''=\mathbf{0},ict''=0)$ of \textbf{MS}$_{0}$. Also, $\mathbf{r}'' \equiv (x'',y'',z'')$.

\item $\mathbf{E}''(x'',t'')$ denotes the electric field, $\mathbf{B}''(x'',t'')$ the magnetic field, $\mathbf{P}''(x'',t'')$ the polarization, $\mathbf{M}''(x'',t'')$ the magnetization, $\mathbf{J}_{f}''(x'',t'')$ the free current density, and $\rho_{f}''(x'',t'')$ the free charge density.
\end{enumerate}
\noindent
Using the Lorentz transformation in (\ref{7}) one can relate the coordinates of the midpoint $q'(t')$ in \textbf{LS}$_{0}$ with those of the midpoint $q''(t'')$ of the mirror along the $x''$-axis in \textbf{MS}$_{0}$. Consider the event whose coordinates in \textbf{LS}$_{0}$ are given by
\begin{eqnarray}
\label{EventsLS0}
b' &=& \left( q'(t'), y', z', ict' \right) \ .
\end{eqnarray}
From (\ref{3}) it is clear that $b'$ is an event associated with a midpoint of the mirror. Using (\ref{7}) it follows that the aforementioned event has coordinates in \textbf{MS}$_{0}$ given by
\begin{eqnarray}
\label{EventsMS0}
b'' &=& \left( q''(t''), y', z', ict'' \right) \ ,
\end{eqnarray}
with
\begin{eqnarray}
\label{9p2}
q''(t'') &=& \gamma_{0}\left[ q'(t') - \beta_{0}ct' \right] \ , \cr
ct'' &=& \gamma_{0}\left[ ct' -\beta_{0}q'(t') \right] \ .
\end{eqnarray}
Using (\ref{midpointLS0}) and (\ref{7}) one can also relate the velocities and accelerations in \textbf{LS}$_{0}$ with those in \textbf{MS}$_{0}$. In particular, for the midpoint one has
\begin{eqnarray}
\label{9p4}
\frac{dq''}{dt''}(t'') &=& \frac{\dot{q}(t'+t_{0}) - v_{0}}{1 - \frac{\beta_{0}}{c}\dot{q}(t'+t_{0})} \ , \cr
\frac{d^{2}q''}{dt''^{2}}(t'') &=& \frac{\ddot{q}(t'+t_{0})}{\gamma_{0}^{3}\left[ 1 - \frac{\beta_{0}}{c}\dot{q}(t'+t_{0}) \right]^{3}} \ .
\end{eqnarray}
From (\ref{midpointLS00}), (\ref{5}), (\ref{9p2}), and (\ref{9p4}) one obtains at time $t'=0$ that 
\begin{eqnarray}
\label{10}
t'' &=& 0 \ , \ \ \ \frac{dq''}{dt''}(0) \ = \ 0 \ , \cr
q''(0) &=& 0 \ , \ \ \ \frac{d^{2}q''}{dt''^{2}}(0) \ = \ \gamma_{0}^{3}\ddot{q}(t_{0}) \ .
\end{eqnarray}
Therefore, the midpoint $q''(t'')$ of the mirror is at rest at the coordinate origin in \textbf{MS}$_{0}$ at time $t''=0$, although it can have a non-zero acceleration. Note that, in general, the argument above does not imply that the other points of the mirror are at rest in \textbf{MS}$_{0}$ at time $t'=0$ because the events with coordinates $(x'\not= 0,t'=0)$ in \textbf{LS}$_{0}$ have coordinates $(x''\not= 0, t''\not=0 )$ in \textbf{MS}$_{0}$. This is related to the issue of rigid bodies in special relativity \cite{Griffiths}.

In the following we assume that \textbf{MS}$_{0}$ is an inertial reference frame in which all the points of the mirror are instantaneously at rest at time $t''=0$. Notice that this assumption holds only approximately if the mirror does not move with constant velocity. 

It follows that the mirror occupies the following region in \textbf{MS}$_{0}$ at time $t''=0$:
\begin{eqnarray}
\label{Region}
R''(0) &=& \left\{ \mathbf{r}''\in\mathsf{R}^{3}: \ \left\vert x'' \right\vert \leq  \frac{\delta_{0}}{2} \right\} \ .
\end{eqnarray}
Recall that $\delta_{0}$ is the thickness of the mirror along the $x''$-axis when it is at rest. 

Since there is vacuum outside the mirror, we know that the polarization, magnetization, and free current and charge densities are zero in \textbf{LS} at time $t_{0}$ for all $x$ outside the mirror, that is, for all $x$ such that $\vert x -q(t_{0}) \vert > \delta(t_{0})/2$, see (\ref{1}). Therefore, we only have to determine these quantities in \textbf{LS} at time $t_{0}$ for all $x$ inside the mirror, that is, for all $x$ such that $\vert x -q(t_{0}) \vert \leq \delta(t_{0})/2$. Equivalently, we have to determine them in \textbf{LS}$_{0}$ at time $t'=0$ for all $\vert x' \vert \leq \delta(t_{0})/2$, see (\ref{4}). We want to take advantage of the rest properties of the mirror. Hence, we can determine the various quantities first in \textbf{MS}$_{0}$ where the mirror is instantaneously at rest at time $t''=0$ and then transform them appropriately to \textbf{LS}$_{0}$. Nevertheless, the condition that the mirror is instantaneously at rest at time $t''=0$ is not enough to be able to use the rest properties to calculate the required quantities in \textbf{LS}$_{0}$ at time $t'=0$. The reason for this is that, as mentioned above, according to (\ref{7}) the events with coordinates $(x'\not= 0,t'=0)$ in \textbf{LS}$_{0}$ have coordinates $(x''\not=0,t''\not=0)$ in \textbf{MS}$_{0}$ and all the points of the mirror may not be at rest in \textbf{MS}$_{0}$ for $t''\not= 0$ (for example, if the mirror does not move with constant velocity in \textbf{LS}, then, in general, not all of its points will be at rest in \textbf{MS}$_{0}$ for $t'' \not=0$). Consequently, one has to assume that the mirror is approximately at rest in \textbf{MS}$_{0}$ during a small time interval centred at $t''=0$ to be able to use the mirror's rest properties. We now determine this time interval explicitly.

First observe from (\ref{7}) the following relation: 
\begin{eqnarray}
\label{implicacion}
&& x' \in \left[ -\frac{\delta(t_{0})}{2}, \ \frac{\delta(t_{0})}{2}  \right], \ \ t'=0 \cr
&\Rightarrow& 
\left\{
\begin{array}{cc}
x'' \in \left[ -\gamma_{0}\frac{\delta(t_{0})}{2}, \gamma_{0}\frac{\delta(t_{0})}{2}  \right], \cr
t'' \in \left[ -\gamma_{0}\vert\beta_{0}\vert\frac{\delta(t_{0})}{2c}, \gamma_{0}\vert\beta_{0}\vert\frac{\delta(t_{0})}{2c}  \right] \ .
\end{array}
\right.
\end{eqnarray}
In other words, points $x'$ in the mirror at time $t'=0$ in \textbf{LS}$_{0}$ correspond in \textbf{MS}$_{0}$ to points $x''$ in the mirror at some time $t''$ in the interval on the right-hand side of (\ref{implicacion}). If we assume that the mirror is at rest in \textbf{MS}$_{0}$ for all $t''$ in the aforementioned interval, then we will be able to use the \textit{rest properties} of the mirror for $(x'',t'')$ corresponding to $(x',t')$ such that $x'$ is inside the mirror and $t'=0$.  

With the discussion of the two previous paragraphs in mind we make the stronger assumption that the mirror is (approximately) at rest in \textbf{MS}$_{0}$ during the time interval
\begin{eqnarray}
\label{N12}
\left[ \ -t_{1}'' , \ t_{1}''  \ \right] \ , \ \ t_{1}'' = \gamma_{0}\vert\beta_{0}\vert\frac{\delta(t_{0})}{2c} \ .
\end{eqnarray}
The mirror is approximately at rest in \textbf{MS}$_{0}$ during the time interval given in (\ref{N12}) if and only if it is subject to very small accelerations in \textbf{MS}$_{0}$. This happens if and only if the magnitudes of the electric and magnetic fields are not very large, since the force affecting the mirror depends on the fields (see Sec. VI). In addition, the results of Sec. VI indicate that these requirements are satisfied in most experimental situations. Also, we consider that the mirror is approximately at rest in \textbf{MS}$_{0}$ during the time interval given in (\ref{N12}) if the midpoint $q''(t'')$ moves a distance much smaller than $\delta_{0}/2$ during this time interval. Recall that $\delta_{0}$ is the thickness of the mirror along the $x''$-axis when it is at rest.

Since (by assumption) the mirror is (approximately) at rest in \textbf{MS}$_{0}$ for $t''\in[-t_{1}'',t_{1}'']$, it follows that the region occupied by the mirror in \textbf{MS}$_{0}$ for $t''\in[-t_{1}'',t_{1}'']$ is given by
\begin{eqnarray}
\label{Region2}
R''(t'') &=& R''(0) \ = \ \left\{ \mathbf{r}''\in\mathsf{R}^{3}: \ \left\vert x'' \right\vert \leq  \frac{\delta_{0}}{2} \right\} \ .
\end{eqnarray}
Notice that the $x''$ interval in (\ref{Region2}) must coincide with the $x''$ interval in (\ref{implicacion}) because they both correspond to the region occupied by the mirror in \textbf{MS}$_{0}$ along the $x''$-axis and the mirror is (approximately) at rest in \textbf{MS}$_{0}$ for $t''\in[-t_{1}'',t_{1}'']$. Therefore, 
\begin{eqnarray}
\label{RelacionDeltas}
\delta(t_{0}) &=& \frac{\delta_{0}}{\gamma_{0}} \ .
\end{eqnarray}
Notice that (\ref{RelacionDeltas}) states that the mirror appears to be Lorentz contracted along the $x$-axis to an observer in \textbf{LS}. 

Finally recall that the electric susceptibility $\chi$ and the conductivity $\sigma$ of the mirror are continuously differentiable functions that are zero outside of the mirror when it is at rest. From (\ref{Region2}) it follows that 
\begin{eqnarray}
\label{chiS}
\chi (x'') = 0 \ , \ \ \sigma(x'') = 0 \ \ \mbox{for} \ x'' \not\in \left( -\frac{\delta_{0}}{2} , \frac{\delta_{0}}{2} \right) .
\end{eqnarray}


\subsection{The electric and magnetic fields}

The electric and magnetic fields can be accommodated into a second-rank anti-symmetric tensor (called the \textit{electromagnetic tensor}) and, therefore, change accordingly under Lorentz transformations \cite{Becker}. One can view these transformation properties in matrix form as follows:
\begin{eqnarray}
\label{VC10}
\left(
\begin{array}{cc}
\mathbf{E}'(x',t') \cr
\mathbf{B}'(x',t')
\end{array}
\right)
&=&
\mathsf{M}_{0}\left(
\begin{array}{cc}
\mathbf{E}''(x'',t'') \cr
\mathbf{B}''(x'',t'')
\end{array}
\right) \ , 
\end{eqnarray}
where 
\begin{eqnarray}
\label{VC9}
\mathsf{M}_{0} &=& 
\left(
\begin{array}{cccccc}
1 & 0 & 0 & 0 & 0 & 0  \cr
0 & \gamma_{0} & 0 & 0 & 0 & \gamma_{0}\beta_{0} \cr
0 & 0 & \gamma_{0} & 0 & -\gamma_{0}\beta_{0} & 0 \cr
0 & 0 & 0 & 1 & 0 & 0 \cr
0 & 0 & -\gamma_{0}\beta_{0} & 0 & \gamma_{0} & 0 \cr
0 & \gamma_{0}\beta_{0} & 0 & 0 & 0 & \gamma_{0}
\end{array}
\right) \ .
\end{eqnarray} 
Here $(x',t')$ are coordinates in \textbf{LS}$_{0}$ and $(x'',t'')$ are the corresponding coordinates in \textbf{MS}$_{0}$ with the connection given by (\ref{7}). Also, the $j$-th component of the column vector on the left of (\ref{VC10}) is $E_{j}'(x',t')$ if $j=1,2,3$ or $B_{j-3}'(x',t')$ if $j=4,5,6$. A similar relation holds for the column vector on the right-hand side of (\ref{VC10}).


\subsection{Polarization and magnetization}

We want to determine the polarization $\mathbf{P}(x,t)$ and magnetization $\mathbf{M}(x,t)$ of the mirror in \textbf{LS} at time $t_{0}$. Equivalently, we can determine $\mathbf{P}'(x',t')$ and $\mathbf{M}'(x',t')$ in \textbf{LS}$_{0}$ at time \ $t'=0$, see (\ref{3p3}). 

The polarization and magnetization can also be accommodated into a second-rank anti-symmetric tensor (called the \textit{moments tensor}) and, therefore, change accordingly under Lorentz transformations \cite{Becker}. The relationship between the polarization and magnetization in \textbf{LS}$_{0}$ and in \textbf{MS}$_{0}$ can also be viewed in matrix form as follows:
\begin{eqnarray}
\label{MatrizPM}
\left(
\begin{array}{cc}
\mathbf{P}'(x',t') \cr
\mathbf{M}'(x',t')
\end{array}
\right)
&=&
\mathsf{M}_{0}^{-1}
\left(
\begin{array}{cc}
\mathbf{P}''(x'',t'') \cr
\mathbf{M}''(x'',t'')
\end{array}
\right) \ .
\end{eqnarray}
Again, $(x',t')$ are coordinates in \textbf{LS}$_{0}$ and $(x'',t'')$ are the corresponding coordinates in \textbf{MS}$_{0}$ with the connection given by (\ref{7}). Also, the $j$-th component of the column vector on the left of (\ref{MatrizPM}) is $P_{j}'(x',t')$ if $j=1,2,3$ or $M_{j-3}'(x',t')$ if $j=4,5,6$. A similar relation holds for the column vector on the right-hand side of (\ref{MatrizPM}). Notice that the electric and magnetic fields are connected by $\mathsf{M}_{0}$, while the polarization and magnetization are connected by $\mathsf{M}_{0}^{-1}$, see (\ref{VC10}) and (\ref{MatrizPM}). This difference is simply due to how the moments and electromagnetic tensors are defined, see the Appendix for the details.

Since (by assumption) the mirror is (approximately) at rest in \textbf{MS}$_{0}$ during the time interval (\ref{N12}) and the mirror is linear, isotropic, and non-magnetizable  when it is at rest, one has
\begin{eqnarray}
\label{iiiN12}
\mathbf{P}''(x'',t'') &=& \chi(x'')\mathbf{E}''(x'',t'') \ , \cr
\mathbf{M}''(x'',t'') &=& \mathbf{0} \qquad \mbox{for all} \ x'' \ \mbox{and} \ t''\in[-t_{1}'',t_{1}'']. \ \
\end{eqnarray}
Recall that the region occupied by the mirror in \textbf{LS}$_{0}$ at time $t'=0$ is given in (\ref{4}). From (\ref{4}), (\ref{implicacion}), (\ref{Region2}), and (\ref{RelacionDeltas}) one has that a point $x'$ inside the mirror at time $t'=0$ in \textbf{LS}$_{0}$ corresponds to a point $x''$ inside the mirror at some time $t''\in[-t_{1}'',t_{1}'']$ in \textbf{MS}$_{0}$. Therefore, (\ref{iiiN12}) is satisfied for the $(x'',t'')$ that corresponds to $x'\in[-\delta(t_{0})/2, \delta(t_{0})/2]$ and $t'=0$. Hence, one can apply (\ref{VC10}), (\ref{MatrizPM}), and (\ref{iiiN12}) to obtain that for $x'\in[-\delta(t_{0})/2, \delta(t_{0})/2]$ and $t'=0$ 
\begin{eqnarray}
\label{T2}
\left(
\begin{array}{cc}
\mathbf{P}'(x',0) \cr
\mathbf{M}'(x',0)
\end{array}
\right)
&=&
\mathsf{M}_{0}^{-1}
\left(
\begin{array}{cc}
\chi(\gamma_{0}x')\mathsf{I}_{3} & \mathsf{O}_{\scriptscriptstyle{3\times 3}} \cr
\mathsf{O}_{\scriptscriptstyle{3\times 3}} & \mathsf{O}_{\scriptscriptstyle{3\times 3}}
\end{array}
\right)
\mathsf{M}_{0}^{-1} \times \cr
&& \qquad \times
\left(
\begin{array}{cc}
\mathbf{E}'(x',0) \cr
\mathbf{B}'(x',0)
\end{array}
\right) \ .
\end{eqnarray}
Here we used (\ref{7}) with $t'=0$ to obtain that $\chi(x'') = \chi(\gamma_{0}x')$. Also, here and in the following $\mathsf{I}_{3}$ is the identity $3\times 3$ matrix and $\mathsf{O}_{n\times m}$ is the $n\times m$ zero matrix.

Also, $\mathbf{P}'(x',0) = \mathbf{M}'(x',0) = \mathbf{0}$ for $\vert x' \vert > \delta(t_{0})/2$ because there is vacuum outside of the mirror, see (\ref{4}). Using (\ref{RelacionDeltas}) and (\ref{chiS}) in (\ref{T2}) one obtains precisely that $\mathbf{P}'(x',0) = \mathbf{M}'(x',0) = \mathbf{0}$ for $\vert x' \vert > \delta(t_{0})/2$. Therefore, (\ref{T2}) is actually valid for all $x'$. Expanding the product in (\ref{T2}) and using (\ref{3p3}) and the fact that $t_{0}$ is arbitrary, one concludes that 
\begin{eqnarray}
\label{14}
\mathbf{P}(x,t) &=& \gamma(t)^{2}\chi_{\mbox{\tiny LS}}(x,t)\left[ \ \mathbf{E}(x,t) + \beta(t)\mathbf{\hat{x}}\times\mathbf{B}(x,t) \right.  \cr
&& \qquad\qquad\qquad \left. -\beta(t)^{2}E_{1}(x,t)\mathbf{\hat{x}} \ \right] \ , \cr
\mathbf{M}(x,t) &=& -\beta(t)\mathbf{\hat{x}} \times \mathbf{P}(x,t) \ ,
\end{eqnarray}
where
\begin{eqnarray}
\label{14p2}
\beta(t) \ \equiv \ \frac{\dot{q}(t)}{c} \ , \qquad \gamma(t) \ \equiv \ \frac{1}{\sqrt{1-\beta(t)^{2}}} \ , 
\end{eqnarray}
and
\begin{eqnarray}
\label{14p3}
\chi_{\mbox{\tiny LS}}(x,t)  \ = \ \chi\left\{ \gamma(t)\left[ x- q(t) \right] \right\} \ .
\end{eqnarray}
Notice that although the mirror is linear, isotropic, and non-magnetizable when it is at rest, to an observer in \textbf{LS} it appears to have a magnetization and the polarization depends not only on the electric field, but also on the velocity of the mirror and on the magnetic field.


\subsection{Free current and charge}

Since the mirror has zero free charge density and satisfies Ohm's law when it is at rest and the mirror is (approximately) at rest during the time interval (\ref{N12}), one has
\begin{eqnarray}
\label{c2}
\rho_{f}''(x'',t'') = 0 \ , \ \ \mathbf{J}_{f}''(x'',t'') = \sigma(x'')\mathbf{E}''(x'',t'') \ ,
\end{eqnarray}
for all $x''$ and $t'' \in [-t_{1}'',t_{1}'']$. From (\ref{c2}) it follows that the current four-vector in \textbf{MS}$_{0}$ is given by \cite{Becker}
\begin{eqnarray}
\label{c3}
\mathbf{s}''(x'',t'') &=& \left( \mathbf{J}_{f}''(x'',t''), \ ic\rho_{f}''(x'',t'') \right)^{T}  \ , \cr
                     &=& \left( \sigma (x'')\mathbf{E}''(x'',t''), \ 0 \right)^{T} \ ,
\end{eqnarray} 
for all $x''$ and \ $t'' \in [-t_{1}'',t_{1}'']$. 

One can express the connection between (\ref{c3}) and the current four-vector $\mathbf{s}'(x',t')$ in \textbf{LS}$_{0}$ \cite{Becker} in matrix form as follows:  
\begin{eqnarray}
\label{c4extra}
\mathbf{s}'(x',t') &=& 
\left( \mathbf{J}_{f}'(x',t'), \ ic\rho_{f}'(x',t') \right)^{T} \ , \cr
&=& \mathsf{M}_{1}\mathbf{s}''(x'',t'') \ ,
\end{eqnarray}
with
\begin{eqnarray}
\label{MatrixM1}
\mathsf{M}_{1} \ = \ 
\left(
\begin{array}{cccc}
\gamma_{0} & 0 & 0 & -i\gamma_{0}\beta_{0} \cr
0 & 1 & 0 & 0 \cr
0 & 0 & 1 & 0 \cr
i\gamma_{0}\beta_{0} & 0 & 0 & \gamma_{0}         
\end{array}
\right) \ .
\end{eqnarray}
Again, $(x',t')$ are coordinates in \textbf{LS}$_{0}$ and $(x'',t'')$ are the corresponding coordinates in \textbf{MS}$_{0}$ with the connection given by (\ref{7}).

Using the assumption that the mirror is at rest during the interval $[-t_{1}'',t_{1}'']$ in (\ref{N12}) and an argument similar to that used with the polarization and magnetization in the previous section (that is, establishing a formula valid for points inside the mirror and then observing that it is also valid for points outside the mirror because there is vacuum), it follows from (\ref{VC10}), (\ref{c3}), and (\ref{c4extra}) that
\begin{eqnarray}
\label{c4}
\mathbf{s}'(x',0) &=& 
\mathsf{M}_{1}
\left(
\begin{array}{cc}
\sigma(\gamma_{0}x')\mathsf{I}_{3} & \mathsf{O}_{\scriptscriptstyle{3\times 3}} \cr                      
\mathsf{O}_{\scriptscriptstyle{1\times 3}} & \mathsf{O}_{\scriptscriptstyle{1\times 3}} 
\end{array}   
\right)
\mathsf{M}_{0}^{-1} \times \cr
&& \qquad \times                   
\left(
\begin{array}{cc}
\mathbf{E}'(x',0) \cr                      
\mathbf{B}'(x',0) 
\end{array}   
\right) \ ,
\end{eqnarray}
for all $x'$. Here we used (\ref{7}) with $t'=0$ to obtain that $\sigma(x'') = \sigma(\gamma_{0}x')$. Expanding the product in (\ref{c4}) and using (\ref{3p3}) along with the fact that $t_{0}$ is arbitrary, one concludes that the free current density $\mathbf{J}_{f}(x,t)$ and the free charge density $\rho_{f}(x,t)$ in \textbf{LS} are given by
\begin{eqnarray}
\label{c5}
\mathbf{J}_{f}(x,t) &=& \gamma(t)\sigma_{\mbox{\tiny LS}}(x,t)\left[ \ \mathbf{E}(x,t) + \beta(t)\mathbf{\hat{x}}\times\mathbf{B}(x,t) \ \right] \ , \cr
\rho_{f}(x,t) &=& \frac{\gamma(t)}{c}\beta(t)\sigma_{\mbox{\tiny LS}}(x,t)E_{1}(x,t) \ ,
\end{eqnarray}
with
\begin{eqnarray}
\label{c5sigma}
\sigma_{\mbox{\tiny LS}}(x,t) &=& \sigma\left\{ \gamma(t)\left[ x- q(t) \right] \right\} \ .
\end{eqnarray}
Notice that, even if the mirror has zero free charge density when it is at rest, it appears to be charged to an observer in \textbf{LS} if $E_{1}(x,t)\not= 0$. Also, observe that the mirror does not satisfy Ohm's law when it is in motion.


\section{MAXWELL'S EQUATIONS}

Maxwell's equations in \textbf{LS} can be written as
\begin{eqnarray}
\label{16}
\nabla \times \mathbf{B}(x,t) &=& \frac{4\pi}{c}\left[ \mathbf{J}_{f}(x,t) + \mathbf{J}_{b}(x,t) \right] + \frac{1}{c}\frac{\partial \mathbf{E}}{\partial t}(x,t) \ ,  \cr
\nabla \times \mathbf{E}(x,t) &=& -\frac{1}{c}\frac{\partial \mathbf{B}}{\partial t}(x,t) \ , \cr
\nabla \cdot \mathbf{E}(x,t) &=& 4\pi \left[ \rho_{f} (x,t) +\rho_{b}(x,t) \right] \ , \cr
\nabla \cdot \mathbf{B}(x,t) &=& 0 \ , 
\end{eqnarray}
where $\rho_{f}(x,t)$ and $\mathbf{J}_{f}(x,t)$ are the free charge and current densities given in (\ref{c5}) and $\rho_{b}(x,t)$ and $\mathbf{J}_{b}(x,t)$ are the bound charge and current. Using (\ref{14}) one has
\begin{eqnarray}
\label{17}
\rho_{b}(x,t) &\equiv& - \nabla\cdot \mathbf{P}(x,t) \ , \cr
&=& -\chi_{\mbox{\tiny LS}}(x,t)\nabla \cdot \mathbf{E}(x,t) - E_{1}(x,t)\frac{\partial}{\partial x} \chi_{\mbox{\tiny LS}}(x,t)  \cr
&&
\end{eqnarray}
and
\begin{eqnarray}
\label{19}
&& \mathbf{J}_{b}(x,t) 
\ \equiv \ \frac{\partial \mathbf{P}}{\partial t}(x,t) + c\nabla \times \mathbf{M}(x,t) \ , \cr
&& \cr
&=& \gamma(t)^{2}\chi_{\mbox{\tiny LS}}(x,t)\left\{ \ \frac{\partial \mathbf{E}}{\partial t}(x,t) \right. \cr
&& \ +\beta(t)\left[ c\frac{\partial\mathbf{E}}{\partial x}(x,t) + \mathbf{\hat{x}}\times \frac{\partial \mathbf{B}}{\partial t}(x,t) -c\frac{\partial E_{1}}{\partial x}(x,t)\mathbf{\hat{x}} \right] \cr
&& \ +\beta(t)^{2}\left[ c \nabla\times\mathbf{B}(x,t)-\frac{\partial\mathbf{E}_{1}}{\partial t}(x,t)\mathbf{\hat{x}}\right] \cr
&& \ \left. + \frac{d\beta}{dt}(t)\left[ \mathbf{\hat{x}}\times\mathbf{B}(x,t) -2\beta(t)E_{1}(x,t)\mathbf{\hat{x}}  \right] \ \right\} \cr
&& \cr
&& + f_{1}(x,t) -\beta(t)f_{2}(x,t) \ .
\end{eqnarray}
Here
\begin{eqnarray}
\label{LasFs}
f_{1}(x,t) &=& \left[ \mathbf{E}(x,t) + \beta (t) \mathbf{\hat{x}} \times \mathbf{B}(x,t) \right]\times \cr
&& \times \left( \beta(t)c\frac{\partial}{\partial x} + \frac{\partial}{\partial t} \right) \gamma(t)^{2}\chi_{\mbox{\tiny LS}}(x,t) \ , \cr
f_{2}(x,t) &=& \mathbf{\hat{x}}E_{1}(x,t)\left( c\frac{\partial}{\partial x} + \beta(t)\frac{\partial}{\partial t} \right) \gamma(t)^{2}\chi_{\mbox{\tiny LS}}(x,t) \ . \cr
&&
\end{eqnarray}
Notice that, when the mirror is at rest (that is, $\dot{q}(t)/c = \beta(t) = 0$  for all $t$), one has $\chi_{\mbox{\tiny LS}}(x,t) = \chi(x-q)$ from (\ref{14p3}) and the right-hand side of (\ref{19}) correctly reduces to $(\partial \mathbf{P}/\partial t)(x,t) = \chi(x-q)(\partial \mathbf{E}/\partial t)(x,t)$ .


\subsection{The case of a piecewise constant susceptibility and conductivity}

Up to now we have assumed that $\chi(x'')$ and $\sigma(x'')$ are continuously differentiable functions. In this subsection we consider the case where they are given by the following two piecewise constant functions:
\begin{eqnarray}
\label{F19}
\chi (x'') &=& 
\left\{
\begin{array}{cc}
\chi_{0} & \mbox{if} \ -\frac{\delta_{0}}{2} \leq  x'' \leq \frac{\delta_{0}}{2} \ , \cr
0 & \mbox{elsewhere}\ . 
\end{array}
\right.
\cr
\sigma (x'') &=& 
\left\{
\begin{array}{cc}
\sigma_{0} & \mbox{if} \ -\frac{\delta_{0}}{2} \leq  x'' \leq \frac{\delta_{0}}{2} \ , \cr
0 & \mbox{elsewhere}\ . 
\end{array}
\right.
\end{eqnarray}
From (\ref{14p3}), (\ref{c5sigma}), and (\ref{F19}) it follows that 
\begin{eqnarray}
\label{F19p2}
\chi_{\mbox{\tiny LS}} (x,t) &=& 
\left\{
\begin{array}{cc}
\chi_{0} & \mbox{if} \ \vert x-q(t) \vert \leq  \frac{\delta(t)}{2} \ , \cr
0 & \mbox{elsewhere}\ . 
\end{array}
\right.
\cr
\sigma_{\mbox{\tiny LS}} (x,t) &=& 
\left\{
\begin{array}{cc}
\sigma_{0} & \mbox{if} \ \vert x -q(t) \vert \leq \frac{\delta(t)}{2} \ , \cr
0 & \mbox{elsewhere}\ . 
\end{array}
\right.
\end{eqnarray}
Here we have used (\ref{RelacionDeltas}) and the fact that $t_{0}\in\mathsf{R}$ is arbitrary to conclude that
\begin{eqnarray}
\label{RelacionDeltas2}
\delta(t) \ = \ \frac{\delta_{0}}{\gamma(t)} \ .
\end{eqnarray} 

All the results we have derived up to now hold with (\ref{F19}). The difference consists in that one must \textit{paste} $\mathbf{E}(x,t)$ and $\mathbf{B}(x,t)$ correctly at the boundaries of the mirror. In other words, one first solves (\ref{16}) inside the mirror, that is, in $R(t)$ given in (\ref{1}). Then, one solves (\ref{16}) outside the mirror, that is, outside of $R(t)$, and, finally, one applies boundary conditions at $x = q(t)\pm \delta(t)/2$. These are obtained by considering the usual boundary conditions for materials at rest in \textbf{MS}$_{0}$ during the time interval $[-t_{1}'',t_{1}'']$ and then using the transformation equations for the field in (\ref{VC10}) to obtain the corresponding boundary conditions in \textbf{LS}$_{0}$. Afterwards, one simply uses (\ref{3p3}) to express the relations with the quantities in \textbf{LS}. This is what we do now. 

Since the mirror is at rest in \textbf{MS}$_{0}$ during the time interval $[-t_{1}'',t_{1}'']$ given in (\ref{N12}), we can use the usual boundary conditions \cite{Jackson}:
\begin{eqnarray}
\label{BC8p1}
\mathbf{B}''\left( x_{0}''+, t'' \right)\cdot\mathbf{\hat{x}}'' &=& \mathbf{B}''\left( x_{0}'' -, t'' \right)\cdot\mathbf{\hat{x}}'' \ , \cr
\mathbf{E}''(x_{0}''+,t'') \times \mathbf{\hat{x}}'' &=& \mathbf{E}''(x_{0}''-,t'') \times \mathbf{\hat{x}}'' \ ,
\end{eqnarray}
and
\begin{eqnarray}
\label{BC8p2}
\left[ \mathbf{D}''\left( x_{0}''+, t'' \right) - \mathbf{D}''\left( x_{0}''-, t'' \right) \right]\cdot \mathbf{\hat{x}}'' &=& 4\pi\sigma_{\scriptscriptstyle{f}}''(x_{0}'',t'')\ , \cr
\mathbf{\hat{x}}'' \times\left[ \mathbf{H}''\left( x_{0}''+, t'' \right) - \mathbf{H}''\left( x_{0}''-, t'' \right) \right] &=& \frac{4\pi}{c}\mathbf{K}_{\scriptscriptstyle{f}}''(x_{0}'',t'')\ , \cr
&&
\end{eqnarray}
for $x_{0}'' = \pm \delta_{0}/2$ and $t''\in[-t_{1}'', t_{1}'']$. Here
\begin{eqnarray}
f(a\pm) \ = \ \mbox{lim}_{x'' \rightarrow a^{\pm}} f(x'') \ .
\end{eqnarray}
In (\ref{BC8p2}) the quantities $\sigma_{f}''(x_{0}'',t'')$ and $\mathbf{K}_{f}''(x_{0}'',t'')$ are the free surface charge and current densities induced at the boundaries of the mirror. Also, from (\ref{iiiN12}) one finds that the electric displacement vector $\mathbf{D}''(x'',t'')$ and the $\mathbf{H}''(x'',t'')$ field are given by 
\begin{eqnarray}
\label{BC8p3}
\mathbf{D}''(x'',t'') &\equiv& \mathbf{E}''(x'',t'') + 4\pi\mathbf{P}''(x'',t'') \ , \cr
                      &=& \left[ 1 + 4\pi\chi (x'') \right]\mathbf{E}''(x'',t'') \ , \cr
                      && \cr
\mathbf{H}''(x'',t'') &\equiv& \mathbf{B}''(x'',t'') - 4\pi\mathbf{M}''(x'',t'') \ , \cr
                      &=& \mathbf{B}''(x'',t'') \ ,
\end{eqnarray}
for all $x''$ and $t''\in [-t_{1}'',t_{1}'']$. 

Since the mirror satisfies Ohm's law and has zero free charge density when it is at rest and the mirror is at rest in \textbf{MS}$_{0}$ during the time interval $[-t_{1}'',t_{1}'']$ given in (\ref{N12}), it follows that \cite{Jackson} 
\begin{eqnarray}
\label{BC8p4}
\mathbf{K}_{\scriptscriptstyle{f}}''(x_{0}'',t'') &=& 0\ , \ \  \sigma_{\scriptscriptstyle{f}}''(x_{0}'',t'') \ = \ 0 \ ,
\end{eqnarray}
for $x_{0}'' = \pm \delta_{0}/2$ and $t''\in[-t_{1}'', t_{1}'']$. 

From (\ref{BC8p1})-(\ref{BC8p4}) it is straightforward to show that
\begin{eqnarray}
\label{BC10}
\mathbf{B}''(x_{0}''+,t'') &=& \mathbf{B}''(x_{0}''-,t'') \ , \cr
E_{j}''(x_{0}''+,t'') &=& E_{j}''(x_{0}''-,t'') \ \ \ \ (j=2,3) \ , \cr
E_{1}''\left( \frac{\delta_{0}}{2}+ ,t'' \right)  
&=&(1+4\pi\chi_{0})E_{1}''\left( \frac{\delta_{0}}{2}- ,t'' \right) \ , \cr
E_{1}''\left( -\frac{\delta_{0}}{2}- ,t'' \right)
&=& (1+4\pi\chi_{0})E_{1}''\left( -\frac{\delta_{0}}{2}+ ,t'' \right) \ ,
\end{eqnarray}
for $x_{0}'' = \pm \delta_{0}/2$ and $t''\in[-t_{1}'', t_{1}'']$. 

From (\ref{7}) and (\ref{RelacionDeltas}) one has that the events with coordinates in \textbf{LS}$_{0}$
\begin{eqnarray}
\left( x_{0}' \ = \ \pm\frac{\delta(t_{0})}{2} \ , \ \ t_{0}' = 0 \right) \ ,
\end{eqnarray} 
have coordinates in \textbf{MS}$_{0}$
\begin{eqnarray}
\left( x_{0}'' \ = \ \pm\delta_{0}/2 \ , \ \ t_{0}'' \ = \ \mp \gamma_{0}\beta_{0}\frac{\delta(t_{0})}{2c} \ = \ \mp \beta_{0}\frac{\delta_{0}}{2c} \right) .
\end{eqnarray}
Hence, it follows from the boundary conditions in (\ref{BC10}) and the relationship between the fields in \textbf{LS}$_{0}$ and in \textbf{MS}$_{0}$ given in (\ref{VC10}) that
\begin{eqnarray}
\label{BC13}
\mathbf{B}'(x_{0}'+,0) &=& \mathbf{B}'(x_{0}'-,0) \ , \cr
E_{j}'(x_{0}'+,0) &=& E_{j}'(x_{0}'-,0) \ \ \ \ (j=2,3) \ , \cr
E_{1}'\left( \frac{\delta(t_{0})}{2}+ ,0 \right) 
&=& (1+4\pi\chi_{0})E_{1}'\left( \frac{\delta(t_{0})}{2}- ,0 \right) \ , \cr
E_{1}'\left( -\frac{\delta(t_{0})}{2}- ,0 \right)
&=& (1+4\pi\chi_{0})E_{1}'\left( -\frac{\delta(t_{0})}{2}+ ,0 \right) , \ \ \ \
\end{eqnarray}
with $x_{0}' = \pm \delta(t_{0})/2$.

Using (\ref{3p3}) to connect the quantities in (\ref{BC13}) with those in \textbf{LS} and recalling that $t_{0}\in\mathbf{R}$ is arbitrary, one concludes that
\begin{eqnarray}
\label{BC}
\mathbf{B}(x_{0}+,t) &=& \mathbf{B}(x_{0}-,t) \ , \cr
E_{j}(x_{0}+,t) &=& E_{j}(x_{0}-,t) \qquad (j=2,3) , \cr
E_{1}(x_{1}-,t) &=& (1+4\pi\chi_{0})E_{1}(x_{1}+,t) , \cr
E_{1}(x_{2}+,t) &=& (1+4\pi\chi_{0})E_{1}(x_{2}-,t), 
\end{eqnarray}
for $x_{0} = q(t) \pm \delta(t)/2$, $x_{1} = q(t) - \delta(t)/2$, $x_{2} = q(t) + \delta(t)/2$, and $t\in\mathsf{R}$. Therefore, the boundary conditions state that the magnetic field $\mathbf{B}(x,t)$ and the tangential components $E_{j}(x,t)$ $(j=2,3)$ of the electric field must be continuous at the boundaries $x = q(t)\pm \delta(t)/2$ of the mirror and that the normal component $E_{1}(x,t)$ has a discontinuity if it is different from zero.


\section{A SPECIAL CASE}

In the rest of the article we assume that the electric field is linearly polarized along the $z$-axis. Then, there is no free and bound charge in \textbf{LS}, see (\ref{c5}) and (\ref{17}), and the electric and magnetic fields can be derived from vector and scalar potentials of the following form:
\begin{eqnarray}
\label{27}
\mathbf{A}(x,t) \ = \ A_{0}(x,t)\mathbf{\hat{z}} \ , \qquad V(x,t) \ = \ 0 \ .
\end{eqnarray}
Explicitly, one has
\begin{eqnarray}
\label{28}
\mathbf{B}(x,t) &=& \nabla \times \mathbf{A}(x,t) \ = \ -\frac{\partial A_{0}}{\partial x}(x,t)\mathbf{\hat{y}} \ , \cr
\mathbf{E}(x,t) &=& -\frac{1}{c}\frac{\partial\mathbf{A}}{\partial t}(x,t) \ = \ -\frac{1}{c}\frac{\partial A_{0}}{\partial t}(x,t)\mathbf{\hat{z}} \ .
\end{eqnarray}
Notice that we are working in the Coulomb gauge and that it coincides with the Lorentz gauge.

With this choice of the electromagnetic field, all the Maxwell equations in (\ref{16}) are automatically satisfied except for the Amp\'{e}re-Maxwell equation (that is, the first equation in (\ref{16})), which now takes the form 
\begin{eqnarray}
\label{E10}
&& \frac{\alpha_{1}(x,t)}{c^{2}}\frac{\partial^{2}A_{0}}{\partial t^{2}}(x,t) + \frac{\alpha_{2}(x,t)}{c}\frac{\partial^{2}A_{0}}{\partial x \partial t}(x,t) \cr
&& \cr
&& + \frac{\alpha_{3}(x,t)}{c}\frac{\partial A_{0}}{\partial x}(x,t) + \frac{\alpha_{4}(x,t)}{c^{2}}\frac{\partial A_{0}}{\partial t}(x,t) \cr
&=& \alpha_{0}(x,t)\frac{\partial^{2}A_{0}}{\partial x^{2}}(x,t) \ .
\end{eqnarray}
Here
\begin{eqnarray}
\label{E9}
\alpha_{0}(x,t) &=& 1 - 4 \pi \gamma (t)^{2} \beta(t)^{2} \chi_{\mbox{\tiny LS}}(x,t) \ , \cr
\alpha_{1}(x,t) &=& 1 + 4 \pi \gamma (t)^{2} \chi_{\mbox{\tiny LS}}(x,t) \ , \cr
\alpha_{2}(x,t) &=& 8\pi \gamma(t)^{2}\beta(t)\chi_{\mbox{\tiny LS}}(x,t) \ , \cr
\alpha_{3}(x,t) &=& 4\pi \frac{d\beta}{dt}(t)\left[ \gamma(t)^{2}\chi_{\mbox{\tiny LS}}(x,t) + \beta(t)f_{0}(x,t) \right] \cr 
&& + 4\pi\gamma(t)\sigma_{\mbox{\tiny LS}}(x,t)\beta(t) \ , \cr
\alpha_{4}(x,t) &=& 4\pi f_{0}(x,t)\frac{d\beta}{dt}(t) +4\pi \gamma(t)\sigma_{\mbox{\tiny LS}}(x,t)\ ,
\end{eqnarray}
and
\begin{eqnarray}
\label{E3}
f_{0}(x,t) &=& \gamma(t)^{4}\beta(t)\left\{ 2\chi(x'') + x'' \frac{d\chi}{dx''}(x'') \right\} , \cr
x'' &=& \gamma(t)[x-q(t)] \ .
\end{eqnarray}
Notice that in (\ref{E3}) it is $\chi$ that appears and not $\chi_{\mbox{\tiny LS}}$, see (\ref{14p3}) for the definition of $\chi_{\mbox{\tiny LS}}$. We remark that (\ref{E10}) was presented at the LAOP 2014 conference \cite{LAOP1} without giving any details of its derivation.

Observe that the coefficients of the partial differential equation in (\ref{E10}) are position- and time-dependent. Therefore, we expect that, in general, the dynamics of the field cannot be restricted to a single-mode, that is, we expect that (\ref{E10}) has no solutions of the form $F(x)T(t)$. 

Up to now we have not taken advantage of two facts: a) one can examine the case where the velocity and acceleration of the mirror are small, and b) the mirror normally evolves on a time-scale much larger than that in which the field evolves. In the following subsection we use these facts to simplify (\ref{E10}).


\subsection{Introduction of non-dimensional quantities}

In the rest of the article we assume that
\begin{enumerate}
\item $\lambda_{0}$ is the characteristic wavelength of the field.
\item $\nu_{0} = c/\lambda_{0}$ is the characteristic frequency of the field.
\item $A_{00}$ is the characteristic value of $A_{0}(x,t)$.
\item $\nu_{\mbox{\tiny osc}}^{-1}$ is the time scale in which $q(t)$ changes appreciably.
\end{enumerate}
We measure length in units of $\lambda_{0}$ and time in units of $\nu_{0}^{-1}$, that is, we take \ $x = \lambda_{0}\xi$ \ and \ $t = \nu_{0}^{-1}\tau = \nu_{\mbox{\tiny osc}}^{-1}\tau_{\mbox{\tiny osc}}$.

Define
\begin{eqnarray}
\label{E6}
\epsilon_{\mbox{\tiny pert}} &=&\frac{\nu_{\mbox{\tiny osc}}}{\nu_{0}} \ , \qquad \ \tilde{\tilde{q}}(\tau_{\mbox{\tiny osc}}) \ = \ \frac{q(\nu_{\mbox{\tiny osc}}^{-1}\tau_{\mbox{\tiny osc}})}{\lambda_{0}} \ , \cr
\tilde{\chi}(\xi) &=& \chi(\lambda_{0}\xi) \ , \qquad \tilde{q}(\tau) \ = \ \frac{q(\nu_{0}^{-1}\tau)}{\lambda_{0}}\ , \cr
\tilde{\sigma}(\xi) &=& \frac{\sigma(\lambda_{0}\xi)}{\nu_{0}} \ , \qquad \ \ \ \tilde{\delta}_{0} = \frac{\delta_{0}}{\lambda_{0}} \ , \cr
\tilde{A}_{0}(\xi , \tau) &=& \frac{1}{A_{00}}A_{0}(\lambda_{0}\xi , \nu_{0}^{-1}\tau) \ .
\end{eqnarray}
Notice that these are non-dimensional quantities and that $\epsilon_{\mbox{\tiny pert}}$ compares the time scale $\nu_{0}^{-1}$ in which the field changes appreciably with the time-scale $\nu_{\mbox{\tiny osc}}^{-1}$ in which $q(t)$ changes appreciably. Since one normally has $\nu_{0}^{-1} \ll \nu_{\mbox{\tiny osc}}^{-1}$ (that is, the field evolves on a much smaller time-scale than the mirror), one expects that $\epsilon_{\mbox{\tiny pert}} \ll 1$. For example, using the experimental values from \cite{NatureMM} one has $\nu_{0} = 2.82 \times 10^{14}$ Hz, $\nu_{\mbox{\tiny osc}} = 1.34 \times 10^{5}$ Hz, and $\epsilon_{\mbox{\tiny pert}} = 4.8 \times 10^{-10}$. Moreover, observe that
\begin{eqnarray}
\label{RelacionQs}
\tilde{q}(\tau) \ = \ \tilde{\tilde{q}}(\epsilon_{\mbox{\tiny pert}}\tau) \ .
\end{eqnarray}

A straightforward calculation using the quantities defined in (\ref{E6}) shows that the non-dimensional form of (\ref{E10}) is given by 
\begin{eqnarray}
\label{E11}
&& \alpha_{1}(x,t)\frac{\partial^{2}\tilde{A}_{0}}{\partial \tau^{2}}(\xi , \tau) \ + \ \alpha_{2}(x,t)\frac{\partial^{2}\tilde{A}_{0}}{\partial \xi \partial \tau}(\xi , \tau)  \cr
&& \cr
&& + \ \frac{\alpha_{3}(x,t)}{\nu_{0}}\frac{\partial \tilde{A}_{0}}{\partial \xi}(\xi , \tau) \ + \ \frac{\alpha_{4}(x,t)}{\nu_{0}}\frac{\partial \tilde{A}_{0}}{\partial \tau}(\xi ,\tau) \cr
&=& \alpha_{0}(x,t)\frac{\partial^{2}\tilde{A}_{0}}{\partial \xi^{2}}(\xi , \tau) \ ,
\end{eqnarray}
with \ $x = \lambda_{0}\xi$ \ and \ $t = \nu_{0}^{-1}\tau$.

In the following we deduce an approximate equation for $\tilde{A}_{0}(\xi,\tau)$ for the case in which the velocity and the acceleration of the mirror are small. In order to do this, assume that 
\begin{eqnarray}
\label{E17p1}
\frac{d\tilde{q}}{d\tau}(\tau) &=& \epsilon_{\mbox{\tiny P}}\tilde{Q}(\tau) \ ,
\end{eqnarray}
where $0 < \epsilon_{\mbox{\tiny P}} \ll 1$ is a perturbation parameter. Notice that \ $\epsilon_{\mbox{\tiny P}} \not= \epsilon_{\mbox{\tiny pert}}$. Recall that $\epsilon_{\mbox{\tiny pert}}$ compares the time-scale in which the field changes appreciably with the time-scale in which $q(t)$ changes appreciably, see (\ref{E6}). Equation (\ref{E17p1}) serves as a definition of $\tilde{Q}(\tau)$. We are simply factoring out $\epsilon_{\mbox{\tiny P}}$ from $(d\tilde{q}/d\tau)(\tau)$, so that the perturbation parameter appears explicitly. 

Before proceeding we delve on the meaning of (\ref{E17p1}). From (\ref{14p2}) and (\ref{E6}) one has 
\begin{eqnarray}
\label{RelacionBetaV}
\frac{\dot{q}(\nu_{0}^{-1}\tau)}{c} \ = \ \beta(\nu_{0}^{-1}\tau) \ = \ \frac{d\tilde{q}}{d\tau}(\tau) \ ,
\end{eqnarray}
so that $\epsilon_{\mbox{\tiny P}}$ in (\ref{E17p1}) is a perturbation parameter that indicates that the velocity of the mirror is very small compared to the speed of light $c$. From (\ref{E17p1}) it follows that
\begin{eqnarray}
\label{E17p2}
\frac{d^{2}\tilde{q}}{d\tau^{2}}(\tau) &=& \epsilon_{\mbox{\tiny P}}\frac{d\tilde{Q}}{d\tau}(\tau) \ .
\end{eqnarray}
Notice that (\ref{RelacionBetaV}) implies that 
\begin{eqnarray}
\label{RelacionBetaV2}
\frac{\ddot{q}(\nu_{0}^{-1}\tau)}{c\nu_{0}} \ = \ \frac{1}{\nu_{0}}\frac{d\beta}{dt}(\nu_{0}^{-1}\tau) \ = \ \frac{d^{2}\tilde{q}}{d\tau^{2}}(\tau) \ ,
\end{eqnarray}
so that $\epsilon_{\mbox{\tiny P}}$ in (\ref{E17p2}) is a perturbation parameter that also indicates that the acceleration of the mirror is very small compared to the speed of light $c$ multiplied by the characteristic frequency $\nu_{0}$ of the field. From the discussion above it follows that an approximation to first order in $\epsilon_{\mbox{\tiny P}}$ corresponds to an approximation to first order in the velocity $\dot{q}(t)$ and the acceleration $\ddot{q}(t)$ of the mirror.

Using (\ref{E6}), (\ref{E17p1}), and (\ref{E17p2}) and neglecting terms of order $\epsilon_{\mbox{\tiny P}}^{n}$ with \ $n\geq 2$ \ it follows from (\ref{E11}) that
\begin{eqnarray}
\label{E21}
\frac{\partial^{2}\tilde{A}_{0}}{\partial \xi^{2}}(\xi , \tau)
&=& \left\{ 1 + 4\pi\tilde{\chi}\left[ \xi - \tilde{q}(\tau) \right] \right\} \frac{\partial^{2}\tilde{A}_{0}}{\partial \tau^{2}}(\xi , \tau)  \cr
&& +4\pi\tilde{\sigma}\left[ \xi -\tilde{q}(\tau)\right] \frac{\partial\tilde{A}_{0}}{\partial \tau}(\xi , \tau) \cr
&& + 8\pi\tilde{\chi}\left[ \xi - \tilde{q}(\tau) \right] \frac{d\tilde{q}}{d\tau}(\tau) \frac{\partial^{2}\tilde{A}_{0}}{\partial \xi\partial \tau}(\xi , \tau) \ , \cr
&& + 4\pi\tilde{\sigma}\left[\xi -\tilde{q}(\tau)\right]\frac{d\tilde{q}}{d\tau}(\tau)\frac{\partial \tilde{A}_{0}}{\partial \xi}(\xi , \tau)   \cr
&& + 4\pi\tilde{\chi}\left[ \xi - \tilde{q}(\tau) \right]\frac{d^{2}\tilde{q}}{d\tau^{2}}(\tau)\frac{\partial \tilde{A}_{0}}{\partial \xi}(\xi , \tau) \ . \cr
&&
\end{eqnarray}
We emphasize that (\ref{E21}) is correct to first order in $\epsilon_{\mbox{\tiny P}}$.

Notice that $\tilde{q}(\tau)$ uses the characteristic time-scale of the field, since $\tau = \nu_{0}t$. It is much better to express equation (\ref{E21}) in terms of the non-dimensional position of the mirror $\tilde{\tilde{q}}(\tau_{\mbox{\tiny osc}})$ because it uses the time-scale of the mirror $\tau_{\mbox{\tiny osc}} = \nu_{\mbox{\tiny osc}}t = \epsilon_{\mbox{\tiny pert}}\tau$ and this allows two of the different time-scales involved in the system to appear explicitly in terms of the perturbation parameter $\epsilon_{\mbox{\tiny pert}}$. Using the relationship \ $\tilde{q}(\tau) = \tilde{\tilde{q}}(\epsilon_{\mbox{\tiny pert}}\tau)$ \ given in (\ref{RelacionQs}), it follows from (\ref{E21}) that
\begin{eqnarray}
\label{EcField}
&& \left\{ 1 + 4\pi\tilde{\chi}\left[ \xi - \tilde{\tilde{q}}(\epsilon_{\mbox{\tiny pert}}\tau) \right] \right\} \frac{\partial^{2}\tilde{A}_{0}}{\partial \tau^{2}}(\xi , \tau)  \cr
&& +4\pi\tilde{\sigma}\left[ \xi -\tilde{\tilde{q}}(\epsilon_{\mbox{\tiny pert}}\tau)\right] \frac{\partial\tilde{A}_{0}}{\partial \tau}(\xi , \tau) \cr
&& + \epsilon_{\mbox{\tiny pert}}8\pi\tilde{\chi}\left[ \xi - \tilde{\tilde{q}}(\epsilon_{\mbox{\tiny pert}}\tau) \right] \frac{d\tilde{\tilde{q}}}{d\tau_{\mbox{\tiny osc}}}(\epsilon_{\mbox{\tiny pert}}\tau) \frac{\partial^{2}\tilde{A}_{0}}{\partial \xi\partial \tau}(\xi , \tau) \cr
&& + \epsilon_{\mbox{\tiny pert}}4\pi\tilde{\sigma}\left[\xi -\tilde{\tilde{q}}(\epsilon_{\mbox{\tiny pert}}\tau)\right]\frac{d\tilde{\tilde{q}}}{d\tau_{\mbox{\tiny osc}}}(\epsilon_{\mbox{\tiny pert}}\tau)\frac{\partial \tilde{A}_{0}}{\partial \xi}(\xi , \tau)  \cr
&& + \epsilon_{\mbox{\tiny pert}}^{2}4\pi\tilde{\chi}\left[ \xi - \tilde{\tilde{q}}(\epsilon_{\mbox{\tiny pert}}\tau) \right]\frac{d^{2}\tilde{\tilde{q}}}{d\tau_{\mbox{\tiny osc}}^{2}}(\epsilon_{\mbox{\tiny pert}}\tau)\frac{\partial \tilde{A}_{0}}{\partial \xi}(\xi , \tau) \cr 
&=& \frac{\partial^{2}\tilde{A}_{0}}{\partial \xi^{2}}(\xi , \tau) \ .
\end{eqnarray}
We emphasize again that (\ref{EcField}) is correct to first order in $\epsilon_{\mbox{\tiny P}}$. Notice that (\ref{EcField}) is a wave equation with damping and with slowly varying coefficients plus terms multiplied by $\epsilon_{\mbox{\tiny pert}}$ and $\epsilon_{\mbox{\tiny pert}}^{2}$ that are a small perturbation when \ $ \epsilon_{\mbox{\tiny pert}} \ll 1$ \ (which is the usual case). Therefore, a multiple-scales approach \cite{Holmes} is an adequate perturbation method to solve it approximately. This is the subject of future work \cite{MultipleScalesField}.

For completeness we now express (\ref{E21}) with units. Using (\ref{E6}) it follows that
\begin{eqnarray}
\label{EcFieldUnits}
\frac{\partial^{2}A_{0}}{\partial x^{2}}(x , t) &=& 
\frac{ 1 + 4\pi\chi\left[ x - q(t) \right] }{c^{2}} \frac{\partial^{2}A_{0}}{\partial t^{2}}(x , t)  \cr
&& +4\pi\frac{\sigma\left[ x -q(t)\right]}{c^{2}} \frac{\partial A_{0}}{\partial t}(x , t) \cr
&& + 8\pi \frac{\chi \left[ x - q(t) \right]}{c^{2}} \dot{q}(t) \frac{\partial^{2}A_{0}}{\partial x \partial t}(x , t)  \cr
&& + 4\pi\frac{\sigma\left[ x -q(t) \right]}{c^{2}}\dot{q}(t)\frac{\partial A_{0}}{\partial x}(x , t)   \cr
&& + 4\pi\frac{\chi\left[ x - q(t) \right]}{c^{2}}\ddot{q}(t)\frac{\partial A_{0}}{\partial x}(x , t) \ . \ \ \ \
\end{eqnarray}
Notice that this equation is correct to first order in $\dot{q}(t)$ and $\ddot{q}(t)$ because it comes from a non-dimensional equation that is correct to first order in $\epsilon_{\mbox{\tiny P}}$.

We note that, taking $\sigma = 0$ and $\chi$ in (\ref{F19}) and using appropriate units, (\ref{EcFieldUnits}) coincides with equation (7) of \cite{Law2} (see also references cited therein). That reference obtained it using an approximate Lagrangian density correct to first order in $\beta(t)$ for the field, see also Sec. VII below.


\subsection{The case of piecewise constant susceptibility and conductivity}

If the electric susceptibility $\chi(x'')$ and the conductivity $\sigma(x'')$ are given by (\ref{F19}), then one solves (\ref{E10}), (\ref{E11}), (\ref{E21}), (\ref{EcField}), and (\ref{EcFieldUnits}) inside and outside of the mirror and one then pastes them using the boundary conditions in (\ref{BC}). With the choice of $\mathbf{A}(x,t)$ given in (\ref{27}), the latter simply amounts to asking that $A_{0}(x,t)$, $(\partial A_{0}/\partial t)(x,t)$, and $(\partial A_{0}/\partial x)(x,t)$ be continuous at the boundaries of the mirror or, equivalently, that $\tilde{A}_{0}(\xi,\tau)$, $(\partial \tilde{A}_{0}/\partial \tau)(\xi,\tau)$, and $(\partial \tilde{A}_{0}/\partial \xi)(\xi,\tau)$ be continuous at the boundaries of the mirror. These are located at $x= q(t)\pm \delta(t)/2$ if (\ref{E10}) or (\ref{E11}) are used. If (\ref{E21}), (\ref{EcField}), or (\ref{EcFieldUnits}) are used, then they are located at $x= q(t)\pm \delta_{0}/2$ because one has to make an approximation to first order in $\epsilon_{\mbox{\tiny P}}$, see (\ref{RelacionDeltas2}). 


\section{FORCE ON THE MIRROR}

In equations (\ref{E10}), (\ref{E11}), (\ref{E21}), (\ref{EcField}), and (\ref{EcFieldUnits}) governing the dynamics of the field, the mirror could have a position $q(t)$ determined by an external agent. In this section we determine the force that the field exerts on the mirror and, consequently, the equation governing its dynamics, namely, the time-derivative of the mechanical momentum (per unit area) equals the force (per unit area). One of the aforementioned equations combined with one for the mirror constitutes a self-consistent set of equations governing the dynamics of the field-mirror system.

Consider the volume
\begin{eqnarray}
\label{F2}
V(t) &=& \left[ q(t)-\frac{\delta (t)}{2} , \ q(t)+\frac{\delta (t)}{2} \right]\times [y_{0}, y_{1}] \times [z_{0}, z_{1}] \ , \cr
&&
\end{eqnarray}
with $y_{0} < y_{1}$ and $z_{0} < z_{1}$. 

Recall that there is no free and bound charge with the choice of $\mathbf{A}(x,t)$ in (\ref{27}), see (\ref{c5}), (\ref{17}), and (\ref{28}). Therefore, the force on the mirror in $V(t)$ at time $t$ is given by \cite{Becker, Jackson}
\begin{eqnarray}
\label{F11}
\mathbf{F}(t) &=& \int_{V(t)}d^{3}r \frac{1}{c}\mathbf{J}(\mathbf{r},t)\times\mathbf{B}(\mathbf{r},t) \ .
\end{eqnarray}

One can rewrite $\mathbf{F}(t)$ using the density of electromagnetic momentum $\mathbf{g}_{\mbox{\tiny em}}(x,t)$ and the Maxwell stress tensor $\mathsf{T}(x,t)$ \cite{Becker,Jackson}. Recall that they are given by
\begin{eqnarray}
\label{F12p2}
\mathbf{g}_{\mbox{\tiny em}}(x,t) &=& \frac{1}{4\pi c}\mathbf{E}(x,t)\times \mathbf{B}(x,t) \ , \cr
\mathsf{T}(x,t) &=& \frac{1}{4\pi}\left[ \mathbf{E}(x,t)\mathbf{E}(x,t)^{T} - \frac{1}{2}\mathbf{E}(x,t)^{2}\mathsf{I}_{3} \right] \cr
&& + \frac{1}{4\pi}\left[ \mathbf{B}(x,t)\mathbf{B}(x,t)^{T} - \frac{1}{2}\mathbf{B}(x,t)^{2}\mathsf{I}_{3} \right] . \ \ \ \
\end{eqnarray}
Here $\mathbf{E}(x,t)$ and $\mathbf{B}(x,t)$ are taken to be column vectors. Using (\ref{F12p2}) it follows that \cite{Jackson}
\begin{eqnarray}
\label{F12}
\mathbf{F}(t) &=& \oint_{\partial V(t)} \mathsf{T}(x,t)\mathbf{n}(t) \ da - \int_{V(t)} d^{3}r \frac{\partial}{\partial t}\mathbf{g}_{\mbox{\tiny em}}(x,t) \ . \cr
&&
\end{eqnarray}
Here $\partial V(t)$ is the surface bounding $V(t)$, $\mathbf{n}(t)$ is the unit vector normal to $\partial V(t)$ and exterior to $V(t)$, and $\mathsf{T}(x,t)\mathbf{n}(t)$ is the product of a matrix times a column vector. Notice that one cannot take the partial derivative with respect to $t$ of $\mathbf{g}_{\mbox{\tiny em}}(x,t)$ out of the integral in (\ref{F12}) as in the case where the volume of integration is time-independent.

After a straightforward calculation it follows from (\ref{28}), (\ref{F2}), (\ref{F12p2}), and (\ref{F12}) that
\begin{eqnarray}
\label{F17}
\mathbf{F}(t) &=& \mathbf{\hat{x}}\left\{ \left. -\frac{1}{8\pi}\left[ \frac{\partial A_{0}}{\partial x}(x,t) \right]^{2} \right\vert_{x= q(t)-\delta(t)/2}^{x= q(t)+\delta(t)/2} \right.  \cr
&& \left. + \frac{1}{4\pi c^{2}}\int_{q(t)-\delta(t)/2}^{q(t)+\delta(t)/2} dx \frac{\partial^{2}A_{0}}{\partial t^{2}}(x,t)\frac{\partial A_{0}}{\partial x}(x,t) \right\} \times \cr
&& \times (y_{1}-y_{0})(z_{1}-z_{0}) \ .
\end{eqnarray}
Here and in the following
\begin{eqnarray}
 h(x,t) \Big\vert_{x=a}^{x=b} \ = \ h(b,t) - h(a,t) \ .
\end{eqnarray}

We now relate $\mathbf{F}(t)$ with the mechanical momentum of the mirror. We assume that the mirror has a uniform mass per unit volume $\rho_{\mbox{\tiny M0}}$ when it is at rest. Then its mass density in \textbf{LS} is given by $\rho_{\mbox{\tiny M}}(t) = \gamma(t)\rho_{\mbox{\tiny M$0$}}$. Note that this holds because we are assuming that the mirror is at rest during the time-interval $[-t_{1}'',t_{1}'']$ given in (\ref{N12}). Using (\ref{RelacionDeltas2}) it follows that the amount of mirror mass in $V(t)$ is given by
\begin{eqnarray}
\label{F36}
M &=& \int_{V(t)}\rho_{\mbox{\tiny M}}(t) d^{3}r \ = \ \delta(t)(y_{1}-y_{0})(z_{1}-z_{0})\gamma(t)\rho_{\mbox{\tiny M$0$}}   \cr
&=& (y_{1}-y_{0})(z_{1}-z_{0})M_{0} \ , 
\end{eqnarray}
with $M_{0}$ the mirror's mass per unit area, that is,   
\begin{eqnarray}
\label{F36p2}
M_{0} &=& \rho_{\mbox{\tiny M$0$}}\delta_{0} \ .
\end{eqnarray}

We now consider the mirror in $V(t)$  to be a single point particle with mass $M$ given in (\ref{F36}). Then, its relativistic mechanical momentum is given by \cite{Becker,Griffiths}
\begin{eqnarray}
\label{F36p3}
\mathbf{P}_{\mbox{\tiny mech}}(t) &=& (y_{1}-y_{0})(z_{1}-z_{0})M_{0}\gamma(t)\dot{q}(t)\mathbf{\hat{x}} \ ,
\end{eqnarray}
and the equation of motion of the mirror is given by
\begin{eqnarray}
\label{FUERZAo}
\frac{d}{dt}\mathbf{P}_{\mbox{\tiny mech}}(t) 
\ = \ \mathbf{F}(t) \ .  
\end{eqnarray}
Using (\ref{F17}) and (\ref{F36p3}) the equation above simplifies to
\begin{eqnarray}
\label{FUERZA}
\frac{d}{dt}p(t) &=&  f(t) \ ,
\end{eqnarray}
with $p(t)$ the mechanical momentum of the mirror (along the $x$-axis) per unit area perpendicular to the $x$-axis and $f(t)$ the pressure exerted by the field along the $x$-axis, that is, 
\begin{eqnarray}
\label{FUERZAcomp}
p(t) &=& M_{0}\gamma(t)\dot{q}(t) \cr
f(t) &=& \left. -\frac{1}{8\pi}\left[ \frac{\partial A_{0}}{\partial x}(x,t) \right]^{2} \right\vert_{x= q(t)-\delta(t)/2 }^{x= q(t)+\delta(t)/2 }   \cr
&& + \frac{1}{4\pi c^{2}}\int_{q(t)-\delta(t)/2}^{q(t)+\delta(t)/2 } dx \frac{\partial^{2}A_{0}}{\partial t^{2}}(x,t)\frac{\partial A_{0}}{\partial x}(x,t) \ . \cr
&&
\end{eqnarray}
\textit{Observe that (\ref{FUERZA}) combined with (\ref{E10}) constitutes the self-consistent set of equations governing the dynamics of the mirror-field system.}

In order to further simplify (\ref{FUERZA}) one must use the equation for $A_{0}(x,t)$. We now do this to first order in $\epsilon_{\mbox{\tiny P}}$ or, equivalently, to first order in $\dot{q}(t)$ and $\ddot{q}(t)$. Substituting $\partial^{2}A_{0}(x,t)/\partial t^{2}$ from (\ref{EcFieldUnits}) in $f(t)$ given in  (\ref{FUERZAcomp}) and simplifying one obtains that
\begin{eqnarray}
\label{c25}
M\left[ q(t), t \right] \ddot{q}(t) &=& F_{0}\left[ q(t), t \right] -F_{1}\left[ q(t),t \right]\frac{\dot{q}(t)}{c} \ , \ \ 
\end{eqnarray}
where
\begin{eqnarray}
\label{c25p2}
F_{0} &=& - \frac{1}{2} \int_{q(t) -\frac{\delta_{0}}{2}}^{q(t) + \frac{\delta_{0}}{2}}dx \ \frac{\chi [x-q(t)]}{1+4\pi\chi [x-q(t)]} \times \cr
&& \ \ \ \ \ \ \ \ \ \ \times \frac{\partial}{\partial x}\left[ \frac{\partial A_{0}}{\partial x}(x , t) \right]^{2} \ , \cr
&& \cr
&& - \frac{1}{c^{2}} \int_{q(t) -\frac{\delta_{0}}{2}}^{q(t) + \frac{\delta_{0}}{2}}dx \ \frac{\sigma [x-q(t)]}{1+4\pi\chi [x-q(t)]} \times \cr
&& \ \ \ \ \ \ \ \ \ \ \times \frac{\partial A_{0}}{\partial x}(x , t)\frac{\partial A_{0}}{\partial t}(x , t) \ , \cr
&& \cr
F_{1} &=& \frac{1}{c}\int_{q(t) - \frac{\delta_{0}}{2}}^{q(t) + \frac{\delta_{0}}{2}}dx \left\{  \frac{\sigma [x-q(t)]}{1 + 4\pi\chi [x-q(t)]} \right. \cr
&& \ \ \ \ \ \ \ \left. + \frac{\chi [x-q(t)]}{1 + 4\pi\chi[x-q(t)]}\frac{\partial}{\partial t} \right\} \left[ \frac{\partial A_{0}}{\partial x}(x, t) \right]^{2}  , \cr
&& \cr
M &=& M_{0} + \frac{1}{c^{2}} \int_{q(t) - \frac{\delta_{0}}{2}}^{q(t) + \frac{\delta_{0}}{2}}dx \ \frac{\chi [x-q(t)]}{1 + 4\pi \chi [x-q(t)]} \times \cr
&& \ \ \ \ \ \ \ \times \left[ \frac{\partial A_{0}}{\partial x}(x, t) \right]^{2} \ . \ \
\end{eqnarray}
For simplicity we omitted the point $[q(t),t]$ where $F_{0}, F_{1}$, and $M$ are evaluated in (\ref{c25p2}). We remark that (\ref{c25}) was presented at the LAOP 2014 conference \cite{LAOP1} without giving any details of its derivation.

We emphasize that (\ref{c25}) is correct to first order in $\epsilon_{\mbox{\tiny P}}$ or, equivalently, to first order in $\dot{q}(t)$ and $\ddot{q}(t)$. \textit{Observe that (\ref{EcFieldUnits}) combined with (\ref{c25}) gives a self-consistent set of equations correct to first order in $\epsilon_{\mbox{\tiny P}}$ that governs the dynamics of the mirror-field system}.

Notice that $M\left[ q(t), t \right]$ reduces to $M_{0}$ and the right-hand side of (\ref{c25}) reduces to $F_{0}\left[ q(t), t \right]$ when the mirror is at rest (recall that the time dependent term in $M\left[ q(t), t \right]$ arises from a term linear in the acceleration of the mirror appearing in the force). Therefore, the motion of the mirror and the coupling to the field give rise to two effects. The first one is a position- and time-dependent mass related to the \textit{effective mass} taken in phenomenological treatments of this type of systems \cite{Nuevo2,Mar2,Israel1}. The second one is a velocity-dependent force that can give rise to friction and that is related to the \textit{cooling} of mechanical objects \cite{Nuevo2,Restrepo}. We note that a \textit{friction force} has also been obtained using a scattering matrix approach and a dispersive dielectric constant for a non-conducting \textit{delta-function} mirror in \cite{Xuereb}. We will investigate the dynamics given by (\ref{c25}) in \cite{MultipleScalesMirror}.


\subsection{Piecewise constant susceptibility and conductivity}

The results of Sec. V are valid for arbitrary continuously differentiable $\chi(x'')$ and $\sigma(x'')$ that are zero outside of the mirror, see (\ref{chiS}). One can also consider the case where they are given in (\ref{F19}). In this case, $\chi [x-q(t)] = \chi_{0}$ and $\sigma [x-q(t)] = \sigma_{0}$ for \ $\vert x - q(t) \vert \leq \delta_{0}/2$ to first order in $\epsilon_{\mbox{\tiny P}}$, so that factors involving these quantities can be taken out of the integrals. In particular, the first term in $F_{0}[q(t),t]$ given in (\ref{c25p2}) can be integrated explicitly.

The simplest case for the dynamics of the mirror+field system is to consider a non-conducting material (that is, $\sigma_{0} = 0$) and to make an approximation to order zero in $\dot{q}(t)$ and $\ddot{q}(t)$ in both the equation for the field and in the force affecting the mirror. In this case equations (\ref{EcFieldUnits}) and (\ref{c25}) reduce to 
\begin{eqnarray}
M_{0}\ddot{q}(t) &=& \left. -\frac{1}{2}\left(\frac{\chi_{0}}{1 + 4\pi\chi_{0}}\right)\left[ \frac{\partial A_{0}}{\partial x}(x , t) \right]^{2} \right\vert_{x = q(t) -\frac{\delta_{0}}{2}}^{x = q(t) + \frac{\delta_{0}}{2}} , \nonumber
\end{eqnarray}
\begin{eqnarray}
\label{Orden0}
\frac{\partial^{2}A_{0}}{\partial x^{2}}(x,t) &=& \frac{1 + 4 \pi \chi[x-q(t)]}{c^{2}}\frac{\partial^{2}A_{0}}{\partial t^{2}}(x,t) . \cr
&& 
\end{eqnarray}
Notice that the time-dependent term modifying $M_{0}$ in the last equation of (\ref{c25p2}) disappears in (\ref{Orden0}) because it comes from corrections to the force of first order in $\dot{q}(t)$ and $\ddot{q}(t)$. We remark that (\ref{Orden0}) was presented at the LAOP 2014 conference \cite{LAOP1} without giving any details of its derivation.

The dynamics of (\ref{Orden0}) have already been studied in \cite{NuestroPRA, NuestroIOP}. In these works we assumed that a perfect mirror is fixed at $x=0$ and that the mobile mirror is free to move for $x>0$. Moreover, we assumed that the mobile mirror is very thin so that the electric susceptibility can be approximated by a delta-function: $\chi(x'') = \chi_{00}\delta(x'')$. Among several results, it was found that, within the rotating-wave-approximation, the force on the mirror can be deduced from a periodic potential with period half the wavelength of the field. Finally, we note that the delta-function case $\chi(x'') = \chi_{00}\delta(x'')$ is obtained by taking the following limits in (\ref{Orden0}): $\delta_{0}\downarrow 0$, $\rho_{\mbox{\tiny M0}} \rightarrow + \infty$, and $\chi_{0} \rightarrow +\infty$ so that $\rho_{\mbox{\tiny M0}}\delta_{0} = M_{0} =$ constant and $\delta_{0}\chi_{0} = \chi_{00} = $ constant. 


\section{VALIDITY OF THE ASSUMPTION THAT THE MIRROR IS APPROXIMATELY AT REST}

We have assumed that the mirror is approximately at rest in \textbf{MS}$_{0}$ during the time interval $[-t_{1}'', t_{1}'']$ given in (\ref{N12}). Moreover, we stated that we consider that the mirror is approximately at rest in \textbf{MS}$_{0}$ during the time interval given in (\ref{N12}) if the midpoint $q''(t'')$ in \textbf{MS}$_{0}$ moves a distance much smaller than half the mirror's thickness (when it is at rest) $\delta_{0}/2$ during the time interval. In this section we establish necessary conditions for this to be true in the case of the electric susceptibility in (\ref{F19}), zero conductivity $\sigma(x'') = 0$, and for a field of the form given in (\ref{27}) and (\ref{28}).

Assume that the mirror is at rest in \textbf{MS}$_{0}$ during the time interval $[-t_{1}'', t_{1}'']$ in (\ref{N12}). We now deduce the force per unit area (pressure) acting on the mirror in \textbf{MS}$_{0}$. 

First recall that the potentials can be accommodated into a four-vector as \ $(\mathbf{A}(x,t), iV(x,t))$ \ \cite{Becker}. If one denotes de potentials in \textbf{LS}$_{0}$ by $\mathbf{A}'(x',t')$ and $V'(x',t')$ and the potentials in \textbf{MS}$_{0}$ by $\mathbf{A}''(x'',t'')$ and $V''(x'',t'')$, then it follows that they are connected by the following relation:
\begin{eqnarray}
\label{VectorPotencial}
\left(
\begin{array}{cc}
\mathbf{A}(x,t) \cr
iV(x,t)
\end{array}
\right)
&=&
\left(
\begin{array}{cc}
\mathbf{A}'(x',t') \cr
iV'(x',t')
\end{array}
\right) \ , \cr
&& \cr
&=& \mathsf{M}_{1}
\left(
\begin{array}{cc}
\mathbf{A}''(x'',t'') \cr
iV''(x'',t'')
\end{array}
\right) \ , \cr
&&
\end{eqnarray}
Here $(x,t)$ are coordinates in \textbf{LS}, while $(x',t')$ and $(x'',t'')$ are the corresponding coordinates in \textbf{LS}$_{0}$ and \textbf{MS}$_{0}$, respectively. The connection between the coordinates is given in (\ref{2}) and (\ref{7}). Also, $\mathsf{M}_{1}$ is given in (\ref{MatrixM1}) and notice that (\ref{3p3}) holds with $\mathbf{A}$ and $V$ instead of $f$. Using (\ref{27}) in (\ref{VectorPotencial}) it immediately follows that 
\begin{eqnarray}
\left(
\begin{array}{cc}
\mathbf{A}''(x'',t'') \cr
\cr
iV''(x'',t'')
\end{array}
\right)
&=&
\left(
\begin{array}{cc}
0 \cr
0 \cr
A_{0}(x,t) \cr
0
\end{array}
\right) \ .
\end{eqnarray}
Therefore, potentials of the form (\ref{27}) in \textbf{LS} imply that the fields in \textbf{MS}$_{0}$ can be deduced from potentials $\mathbf{A}''(x'',t'') = A_{0}''(x'',t'')\mathbf{\hat{z}''}$ and $V''(x'',t'') = 0$ with formulae for the fields similar to those in (\ref{28}).

Now observe that the right-hand side of the first equation in (\ref{Orden0}) gives the pressure affecting the mirror when it is at rest, since the right-hand side of that equation gives the correct pressure on mirror to order zero in $\beta (t)$ and $\dot{\beta}(t)$. Therefore, to determine the force acting on the mirror in \textbf{MS}$_{0}$ during the time interval $[-t_{1}'',t_{1}'']$  we simply have to take the right-hand side of the first equation in (\ref{Orden0}) and replace $(\partial A_{0}/\partial x)(x,t)$ by $(\partial A_{0}''/\partial x'')(x'',t'')$ and $q(t)$ by $q''(t'')$ and use that $\mathbf{B}''(x'',t'') = \nabla '' \times \mathbf{A}''(x'',t'') = -(\partial A_{0}''/\partial x'')(x'',t'')\mathbf{\hat{y}}''$. Then, it follows from (\ref{Orden0}) that the pressure affecting the mirror at time $t''=0$ in \textbf{MS}$_{0}$ is given by 
\begin{eqnarray}
\label{N16f3}
f_{0} \ = \ \left. -\frac{\chi_{0}/2}{1 + 4\pi\chi_{0}}\mathbf{B}''(x'',0)^{2}\right\vert_{x''=-\delta_{0}/2}^{x''=\delta_{0}/2} \ .
\end{eqnarray}
Notice that we used $q''(0)=0$, see (\ref{10}). We emphasize that the pressure in (\ref{N16f3}) was obtained assuming that the mirror is at rest in \textbf{MS}$_{0}$ during the time interval $[-t_{1}'',t_{1}'']$. We now determine if the mirror really moves a negligible distance if this pressure is introduced in its equation of motion.

According to (\ref{FUERZA}) the equation of motion of the mirror in \textbf{MS}$_{0}$ is
\begin{eqnarray}
\label{N17}
\frac{d}{dt''}\left[ M_{0}\gamma_{1}(t'')\frac{dq''}{dt''}(t'') \right] &\simeq& f_{0} \ ,
\end{eqnarray}
where 
\begin{eqnarray}
\gamma_{1}(t'') \ = \ \left\{ 1 - \left[ \frac{1}{c}\frac{dq''}{dt''}(t'') \right]^{2} \right\}^{-1/2} \ ,
\end{eqnarray}
and we have approximated the pressure acting on the mirror in \textbf{MS}$_{0}$ at time $t''$ by the pressure acting on it at time $t''=0$.

It is straightforward to show that the solution of (\ref{N17}) with $q''(0)=0$ and $(dq''/dt'')(0) =0$ is given by
\begin{eqnarray}
\label{N18}
q''(t'') \ \simeq \  \frac{f_{0}}{2M_{0}}(t'')^{2} \qquad \mbox{for} \ t''\in[ 0,t_{1}''],
\end{eqnarray}
if 
\begin{eqnarray}
\label{C2}
\left(\frac{f_{0}t_{1}''}{M_{0}c}\right)^{2} \ \ll \ 1 \ .
\end{eqnarray}

It follows from (\ref{N18}) that the mirror will be approximately at rest during the interval $[0, t_{1}'']$ if
\begin{eqnarray}
\label{N19}
\left\vert \frac{f_{0}}{2M_{0}}\left( t_{1}'' \right)^{2} \right\vert &\ll& \frac{\delta_{0}}{2} \ , \qquad \left\vert \frac{f_{0}}{cM_{0}} t_{1}'' \right\vert^{2} \ \ll \ 1 \ .
\end{eqnarray}
The first condition in (\ref{N19}) simply states that the midpoint of the mirror must have a displacement much smaller than half the thickness of the mirror when it is at rest. Meanwhile, the second condition in (\ref{N19}) is identical to (\ref{C2}) and states that the square of the velocity of the midpoint of the mirror must be much smaller than $c^{2}$.

We now obtain a sufficient condition for (\ref{N19}). Assume that
\begin{eqnarray}
\label{N20}
\vert \mathbf{E}(x,t) \vert \leq E_{\mbox{\tiny max}} \ , \qquad \vert \mathbf{B}(x,t) \vert \leq B_{\mbox{\tiny max}} \ .
\end{eqnarray}
Using (\ref{VC10}) in combination with (\ref{28}) it follows that
\begin{eqnarray}
\label{N20p2}
\vert \mathbf{E}''(x'',0) \vert &\leq& \gamma_{0}(E_{\mbox{\tiny max}}+ \vert \beta_{0} \vert B_{\mbox{\tiny max}}) \ , \cr
\vert \mathbf{B}''(x'',0) \vert &\leq& \gamma_{0}(B_{\mbox{\tiny max}}+ \vert \beta_{0} \vert E_{\mbox{\tiny max}}) \ .
\end{eqnarray}
Using (\ref{N20p2}) in (\ref{N19}) it is straightforward to show that (\ref{N19}) will hold whenever
\begin{eqnarray}
\label{N21}
B_{\mbox{\tiny max}}, \ \vert \beta_{0} \vert E_{\mbox{\tiny max}} \ \ll \sqrt{\frac{2\pi \rho_{\mbox{\tiny M0}} c^{2}}{\gamma_{0}^{2}\vert \beta_{0} \vert}} \ .
\end{eqnarray}
We now illustrate (\ref{N21}) with an example. Reference \cite{NatureMM} introduces an experimental set up that can be analyzed with the model of this article. They have a mirror with the following properties: 
\begin{eqnarray}
\label{ParametrosNature}
L &=& W \ = \ 10^{-3} \ \mbox{(m)} \ , \cr
\delta_{0} &=& 50\times 10^{-9} \ \mbox{(m)} , \cr
M &=& 4\times 10^{-11} \ \mbox{(kg)} .
\end{eqnarray}
where $L = (y_{1}-y_{0})$ and $W = (z_{1}-z_{0})$ denote the length and the width of the mirror, respectively. 

The quantity on the right-hand side of (\ref{N21}) is a strictly decreasing function of $\beta_{0}$ and is minimized for $\beta_{0} = 1$, that is, when the speed of the mirror equals the speed of light. With this in mind we take $\beta_{0} = 1/2$, since smaller values of $\beta_{0}$ give larger bounds. 

To relate better to quantities used in a laboratory, we now transform the electric and magnetic fields from Gaussian to MKS units. It can be shown that \cite{Jackson}
\begin{eqnarray}
\label{GaussMKS}
\mathbf{E}^{\mbox{\tiny (MKS)}} &=& \frac{\mathbf{E}^{\mbox{\tiny (gauss)}}}{\sqrt{4\pi\epsilon_{0}}} \ , \ \ \mathbf{B}^{\mbox{\tiny (MKS)}} \ = \ \frac{\mathbf{B}^{\mbox{\tiny (gauss)}}}{c\sqrt{4\pi\epsilon_{0}}} \ ,
\end{eqnarray}
where $\mathbf{E}^{\mbox{\tiny (MKS)}}$ ($\mathbf{B}^{\mbox{\tiny (MKS)}}$) is the electric (magnetic) field in MKS units, $\mathbf{E}^{\mbox{\tiny (gauss)}}$ ($\mathbf{B}^{\mbox{\tiny (gauss)}}$) is the electric (magnetic) field in Gaussian units, and $\epsilon_{0}$ is the permittivity of vacuum. 

From (\ref{N21}), (\ref{GaussMKS}), the parameters in (\ref{ParametrosNature}), and the value $\beta_{0} = 1/2$, it follows that the maximum electric $\mathbf{E}_{\mbox{\tiny max}}^{\mbox{\tiny (MKS)}}$ and magnetic $\mathbf{B}_{\mbox{\tiny max}}^{\mbox{\tiny (MKS)}}$ fields in MKS units must satisfy
\begin{eqnarray}
\label{N212}
E_{\mbox{\tiny max}}^{\mbox{\tiny (MKS)}} &\ll& \frac{1}{\vert \beta_{0}\vert\sqrt{4\pi \epsilon_{0}} }\sqrt{\frac{2\pi \rho_{\mbox{\tiny M0}}c^{2}}{\gamma_{0}^{2}\vert \beta_{0} \vert}} \ , \cr
&\simeq& 5.2 \times 10^{15} \ \left(\frac{\mbox{V}}{\mbox{m}} \right) , \cr
B_{\mbox{\tiny max}}^{\mbox{\tiny (MKS)}} &\ll& \frac{1}{c\sqrt{4\pi\epsilon_{0}}}\sqrt{\frac{2\pi \rho_{\mbox{\tiny M0}}c^{2}}{\gamma_{0}^{2}\vert \beta_{0} \vert}} \ \simeq \ 10^{7} \ \mbox{(T)} .
\end{eqnarray}

To have an idea of the order of magnitude of $E_{\mbox{\tiny max}}^{\mbox{\tiny (MKS)}}$ and $B_{\mbox{\tiny max}}^{\mbox{\tiny (MKS)}}$ we now consider a plane wave describing the electromagnetic field of a laser. It can be shown that the norm of (the average in one cycle of) the Poynting vector of a plane wave in vacuum and in MKS units is given by \cite{Griffiths}
\begin{eqnarray}
\label{Poynting}
\vert \langle \mathbf{S}_{\mbox{\tiny MKS}} \rangle \vert \ = \ \frac{c\epsilon_{0}}{2}\left(E_{\mbox{\tiny max}}^{\mbox{\tiny (MKS)}}\right)^{2} \ .
\end{eqnarray} Here $E_{\mbox{\tiny max}}^{\mbox{\tiny (MKS)}}$ is the magnitude of the electric field of the plane wave in MKS units.
Therefore, the incident power (in Watts) on the mirror is
\begin{eqnarray}
\label{Power}
P &=& LW\vert \langle \mathbf{S}_{\mbox{\tiny MKS}} \rangle \vert \ .
\end{eqnarray}
Factoring out $E_{\mbox{\tiny max}}^{\mbox{\tiny (MKS)}}$, substituting the parameters in (\ref{ParametrosNature}), and using the well-known relationship between the magnitudes of the electric and magnetic fields of a plane wave in vacuum \cite{Griffiths} one gets
\begin{eqnarray}
\label{Eop}
E_{\mbox{\tiny max}}^{(\mbox{\tiny MKS})} &=& 2.7\sqrt{P}\times 10^{4} \ \left( \frac{\mbox{V}}{\mbox{m}} \right) , \cr
B_{\mbox{\tiny max}}^{(\mbox{\tiny MKS})} &=& \frac{1}{c}E_{\mbox{\tiny max}}^{\mbox{\tiny (MKS)}} \ = \ \sqrt{P}\times 10^{-4} \ \left( T \right) .
\end{eqnarray}
For a laser power of $P = 1$ (Watt) one observes that both quantities in (\ref{Eop}) are much smaller than the bounds in (\ref{N212}). Therefore, we conclude that it is a reasonable assumption to consider that the mirror is approximately at rest during the time interval $[-t_{1}'',t_{1}'']$.


\section{A LAGRANGIAN DENSITY FOR THE FIELD}

In this section we derive a Lagrangian density for the field in the case where the conductivity is zero, that is, $\sigma(x'') = 0$. 

Since (\ref{iiiN12}) holds during the time interval $[-t_{1}'',t_{1}'']$, it follows that the Lagrangian density for the electromagnetic field in \textbf{MS}$_{0}$ is given by \cite{Jackson}
\begin{eqnarray}
\label{L1}
\mathcal{L}'' &=& \frac{1}{8\pi}\Big[ \mathbf{E}''(x'',t'')\cdot\mathbf{D}''(x'',t'') \cr
&& \qquad - \mathbf{B}''(x'',t'')\cdot\mathbf{H}''(x'',t'') \Big] \ , \cr
&=& \frac{1}{8\pi}\left[ \epsilon(x'')\mathbf{E}''(x'',t'')^{2} - \mathbf{B}''(x'',t'')^{2} \right] \ ,
\end{eqnarray}
for $t''\in[-t_{1}'',t_{1}'']$ and the dielectric function
\begin{eqnarray}
\label{DielectricFunction}
\epsilon (x'') &=& 1 + 4\pi \chi(x'') \ .
\end{eqnarray}
One can write (\ref{L1}) in matrix form as follows:
\begin{eqnarray}
\label{L2}
\mathcal{L}'' &=& \frac{1}{8\pi}
\left(
\begin{array}{cc}
\mathbf{E}''(x'',t'') \cr
\mathbf{B}''(x'',t'') 
\end{array}
\right)^{T}
\left(
\begin{array}{cc}
\epsilon(x'')\mathsf{I}_{3} & \mathsf{O}_{\scriptscriptstyle{3\times 3}} \cr
\mathsf{O}_{\scriptscriptstyle{3\times 3}} & -\mathsf{I}_{3}
\end{array}
\right) \times \cr
&& \qquad \times
\left(
\begin{array}{cc}
\mathbf{E}''(x'',t'') \cr
\mathbf{B}''(x'',t'') 
\end{array}
\right) \ . 
\end{eqnarray}

Recall that the mirror is at rest in \textbf{MS}$_{0}$ during the time interval $[-t_{1}'',t_{1}'']$ in (\ref{N12}). Then one can use (\ref{VC10}) to connect the fields in \textbf{MS}$_{0}$ with those in \textbf{LS}$_{0}$ to obtain the Lagrangian density in \textbf{LS}$_{0}$ at time $t' = 0$ (here one uses an argument similar to that used to determine the polarization and magnetization where one obtains a formula valid inside the mirror and then observes that it also gives the correct value outside the mirror). If one then uses (\ref{3p3}) to connect the fields in \textbf{LS}$_{0}$  with those in \textbf{LS} and recalls that $t_{0}$ is arbitrary, one concludes that the Lagrangian density in \textbf{LS} is given by
\begin{eqnarray}
\label{L4}
\mathcal{L} &=& \frac{1}{8\pi}
\left(
\begin{array}{cc}
\mathbf{E}(x,t) \cr
\mathbf{B}(x,t) 
\end{array}
\right)^{T} 
\left(\mathsf{M}_{0}^{-1}\right)^{T}
\left(
\begin{array}{cc}
\epsilon(x'')\mathsf{I}_{3} & \mathsf{O}_{\scriptscriptstyle{3\times 3}} \cr
\mathsf{O}_{\scriptscriptstyle{3\times 3}} & -\mathsf{I}_{3}
\end{array}
\right)
\times \cr
&& \times
\mathsf{M}_{0}^{-1}
\left(
\begin{array}{cc}
\mathbf{E}(x,t) \cr
\mathbf{B}(x,t) 
\end{array}
\right) \ ,
\end{eqnarray}
with \ $x'' = \gamma(t)[x-q(t)]$ \ and $\mathsf{A}^{T}$ the transpose of matrix $\mathsf{A}$. 

Equation (\ref{L4}) is valid for an arbitrary electromagnetic field. For the special case given in (\ref{27}) and (\ref{28}) one obtains from (\ref{L4}) the following Lagrangian density:
\begin{eqnarray}
\label{L5}
\mathcal{L} &=& 
\frac{1}{8\pi}
\left\{ 
\left[ \frac{1}{c}\frac{\partial A_{0}}{\partial t}(x,t) \right]^{2}  - \left[ \frac{\partial A_{0}}{\partial x}(x,t) \right]^{2} \right\}  \cr
&& + \frac{\gamma(t)^{2}\chi_{\mbox{\tiny LS}}(x,t)}{2}
\left[ \frac{1}{c}\frac{\partial A_{0}}{\partial t}(x,t) + \beta (t)\frac{\partial A_{0}}{\partial x}(x,t) \right]^{2} . \cr
&&
\end{eqnarray}
Observe that (\ref{L5}) expresses the Lagrangian density as the sum of a part corresponding to the free field plus a part associated with the presence of the mirror. We note that (\ref{L5}) is identical to the Lagrangian density given in \cite{Law2} (see also references therein) for the case of the piecewise constant electric susceptibility $\chi$ given in (\ref{F19}).

The Euler-Lagrange equation associated with $\mathcal{L}$ is then given by
\begin{eqnarray}
\label{L7}
\frac{\partial}{\partial t}\left\{ \frac{\partial \mathcal{L}}{\partial \left[ \partial_{t}A_{0}(x,t)\right]} \right\} + \frac{\partial}{\partial x}\left\{ \frac{\partial \mathcal{L}}{\partial \left[ \partial_{x}A_{0}(x,t)\right]} \right\} &=& 0 \ , \cr
&&
\end{eqnarray}
where $\partial_{t}$ and $\partial_{x}$ denote partial derivatives with respect to $t$ and $x$, respectively. After a lengthy calculation one can show from (\ref{L5}) that the Euler-Lagrange equation (\ref{L7}) does indeed give (\ref{E10}) with $\sigma_{\mbox{\tiny LS}}(x,t) = 0$.

One can be interested in using a simpler Lagrangian density from which approximate equations for the field can be obtained (for example, to first order in the velocity and the acceleration of the mirror). One way to achieve this is to expand the Lagrangian density in (\ref{L5}) in powers of $\beta(t)$ as follows:
\begin{eqnarray}
\label{L15}
\mathcal{L} &=& \mathcal{L}_{0} + \beta(t)\mathcal{L}_{1} + \frac{1}{2}\beta(t)^{2}\mathcal{L}_{2} \ + ... \ , 
\end{eqnarray}
where the first three terms are given by
\begin{eqnarray}
\label{L14}
\mathcal{L}_{0} &=& \frac{1}{8\pi}\left\{ \frac{\epsilon[x-q(t)]}{c^{2}}\left[ \frac{\partial A_{0}}{\partial t}(x,t) \right]^{2} - \left[ \frac{\partial A_{0}}{\partial x}(x,t) \right]^{2} \right\} , \cr
\mathcal{L}_{1} &=&  \chi[x-q(t)]\frac{1}{c}\frac{\partial A_{0}}{\partial t}(x,t)\frac{\partial A_{0}}{\partial x}(x,t)  \ , \cr
\mathcal{L}_{2} &=& \chi[x-q(t)]\left\{ \left[ \frac{1}{c}\frac{\partial A_{0}}{\partial t}(x,t) \right]^{2} + \left[ \frac{\partial A_{0}}{\partial x}(x,t) \right]^{2} \right\} \cr
&&\ \ + \frac{x-q(t)}{2}\frac{d\chi}{dx''}[x-q(t)]\left[ \frac{1}{c}\frac{\partial A_{0}}{\partial t}(x,t) \right]^{2} \ .
\end{eqnarray}
Now we make a few comments on (\ref{L15}). First, the term $\beta(t)^{n}\mathcal{L}_{n}/n!$ introduces terms of order $\epsilon_{\mbox{\tiny P}}^{n}$ and $\epsilon_{\mbox{\tiny P}}^{n+1}$ in the Euler-Lagrange equations for the field, since one has to calculate derivatives with respect to $t$ (see (\ref{L7})) and factors $\chi[x-q(t)]$ are present. Therefore, the equations of order $n$ in $\dot{q}(t)$ and $\ddot{q}(t)$ for the field cannot be deduced exactly with (\ref{L15}) if one neglects $\mathcal{L}_{m}$ for $m>n$, since terms of order $n+1$ would have to be neglected in the Euler-Lagrange equations to obtain the correct equations of motion. This shows that it appears not to be possible to have an approximate Lagrangian density of a given order for the field that yields the correct equations without having to discard terms. The argument above holds for a continuously differentiable electric susceptibility $\chi (x'')$. For the piecewise constant $\chi(x'')$ in (\ref{F19}), the derivatives of $\chi[x-q(t)]$ are zero inside and outside of the mirror, so that one obtains the correct equations of order $n$ in $\dot{q}(t)$ and $\ddot{q}(t)$ for the field if one neglects $\mathcal{L}_{m}$ for $m>n$. Also, in this case one must paste the solution at the boundaries of the mirror so as to have a continuously differentiable function, see (\ref{BC}).

For an approximate Lagrangian of order one in $\beta(t)$ for the complete mirror+field system we refer the reader to \cite{Law2}, where the following Lagrangian is proposed:
\begin{eqnarray}
\label{LLaw2}
L^{(1)} &=& \frac{1}{2}M_{0}\dot{q}(t)^{2} + V[q(t)] + \int_{0}^{L}dx \left[ \mathcal{L}_{0} + \beta(t)\mathcal{L}_{1} \right].  \cr
&&
\end{eqnarray}
Here $V[q(t)]$ is a potential affecting the mirror and $[0, L]$ is the region where the mirror can move. Moreover, two perfect, fixed mirrors are located at $x=0$ and $x=L$. This Lagrangian gives the correct force on the mirror only to order zero in $\dot{q}(t)$ and $\ddot{q}(t)$ and has been used to quantize the mirror+field system \cite{Law2}. Since one does not recover the correct time-dependent mass and velocity-dependent force affecting the mirror, physical phenomena associated with these terms have to be included by other means, such as master equation methods, in a similar way in which damping is introduced in a harmonic oscillator and spontaneous emission is introduced in the interaction of a two level atom with a single-mode electromagnetic field \cite{Nuevo2}.


\section{CONCLUSIONS}

In this article we established the equations that govern the dynamics of a system composed of a mobile slab interacting with the electromagnetic field using a relativistic treatment and not considering thermal effects. The slab is made of a material that satisfies the following properties when it is at rest: it is linear, isotropic, non-magnetizable, and ohmic with zero free charge density. Moreover, we obtained approximate equations for the slab-field system correct to first order in the velocity and the acceleration of the slab. On one hand, we showed that the electromagnetic field satisfies a wave equation with damping and slowly varying coefficients plus terms that are small when the slab evolves on a time-scale much larger than that of the field. These properties arise from the fact that the mobile slab appears to have a magnetization and a polarization that depends on the magnetic field when it is in motion. On the other hand, the slab satisfies a dynamical equation with two terms that arise as a result of the motion of the slab and the coupling to the electromagnetic field. The first one is a position- and time-dependent mass related to the \textit{effective mass} taken in phenomenological treatments of this type of systems. The second is a velocity-dependent force that can give rise to friction and that is related to the \textit{cooling} of mechanical objects. Also, the fact that there are two separate time-scales, one associated with the fast evolution of the field and another associated with the slower evolution of the slab,  allows the use of the multiple scales method \cite{Holmes} to study the dynamics of the system. This is the topic of work in preparation \cite{MultipleScalesField}.

One may inquire why a relativistic treatment is needed, especially in the case where the slab does not move at relativistic speeds. Well, one is confronted with the following problem: one knows the properties of the slab (such as the polarization and the magnetization) only when the slab is at rest and one needs to determine them in the \textit{Laboratory reference frame} \textbf{LS} where it can be in motion. One way to solve this problem is discussed in this article and consists in first calculating the properties of the slab in an inertial reference frame where the slab is approximately at rest during a small time interval and then using their transformation properties under Lorentz transformations to determine their form in \textbf{LS}. The advantage of using Lorentz transformations is threefold. First, all electromagnetic quantities (such as the polarization and the magnetization) have simple, well-defined transformation properties under Lorentz transformations. Second, the use of Lorentz transformations illuminates the results, enables a better understanding of the physics of the system, and allows one to obtain correction terms in the case of small velocities and accelerations. For example, one finds that, although the slab is made up of a linear, isotropic, and non-magnetizable material when it is at rest, when the slab is in motion it has a magnetization and a polarization that depend on the electric and magnetic fields and on its velocity. Moreover, one also finds that these modified magnetization and polarization lead to a time dependent mass and to a velocity-dependent force affecting the motion of the slab. Although these terms are small for small velocities of the slab, they give rise to important phenomena related to the effective mass and \textit{cooling} of mechanical objects. Third, the relativistic treatment allows one to obtain consistent approximations for both the field and the slab at a given order of the velocity and acceleration.

It is also important to note that in our treatment the dynamics of the slab are general and are not restricted, for example, to oscillatory motions. In fact, our treatment allows the slab to accelerate slowly up to relativistic velocities. Moreover, all of our expressions reduce to the correct value when the slab has constant velocity.

Finally, deducing the equations for both the field and the mobile slab constitutes a problem of fundamental physics and these equations can be important in other physical contexts.


\section*{ACKNOWLEDGEMENTS}

R. Weder is a fellow of the Sistema Nacional de Investigadores. L. O. Casta\~{n}os thanks the Universidad Nacional Aut\'{o}noma de M\'{e}xico for support. Research partially supported by project PAPIIT-UNAM IN102215.


\section*{APPENDIX}

In the $ict$-system or Minkowski metric (which is the one used in this article) with coordinates $(x_{1}=x, x_{2}=y, x_{3}=z, x_{4}=ict)$, one first introduces the \textit{electromagnetic tensor} \cite{Becker}:
\begin{eqnarray}
\label{Fnumu}
F_{\nu \mu} &=& \left(
\begin{array}{cccc}
0      & B_{3}  & -B_{2} & -iE_{1} \cr
-B_{3} & 0      &  B_{1} & -iE_{2} \cr
B_{2}  & -B_{1} &  0     & -iE_{3} \cr
iE_{1} & iE_{2} & iE_{3} & 0 
\end{array}
\right) \ .
\end{eqnarray}
Here $E_{j}$ and $B_{j}$ $(j=1,2,3)$ are the (Cartesian) components of the electric $\mathbf{E}$ and magnetic $\mathbf{B}$ fields, respectively.

In order to write Maxwell's equations in four-dimensional form in terms of free charges and free currents only, one has to incorporate the bound charges and bound currents into the fields. This can be accomplished by first introducing the \textit{moments tensor} $M_{\nu \mu}$ given by \cite{Becker}:
\begin{eqnarray}
\label{Mnumu}
M_{\nu \mu} &=& \left(
\begin{array}{cccc}
0      & M_{3}  & -M_{2} &  iP_{1} \cr
-M_{3} & 0      &  M_{1} &  iP_{2} \cr
M_{2}  & -M_{1} &  0     &  iP_{3} \cr
-iP_{1} & -iP_{2} & -iP_{3} & 0 
\end{array}
\right) \ . 
\end{eqnarray}
Here $M_{j}$ and $P_{j}$ $(j=1,2,3)$ are the (Cartesian) components of the magnetization $\mathbf{M}$ and polarization $\mathbf{P}$, respectively. One then defines the tensor $H_{\nu \mu} = (F_{\nu\mu} - 4\pi M_{\nu\mu})$ and makes use of the usual definitions $\mathbf{D} = \mathbf{E} + 4\pi \mathbf{P}$ and $\mathbf{H} = \mathbf{B}-4\pi\mathbf{M}$ to express $H_{\nu \mu}$ in terms of the electric displacement vector $\mathbf{D}$ and the field $\mathbf{H}$.

Using the tensors introduced above one can write Maxwell's equations in four-dimensional form as follows:
\begin{eqnarray}
\label{Forma4D}
\sum_{\mu = 1}^{4}\frac{\partial H_{\nu\mu}}{\partial x_{\mu}} &=& \frac{4\pi}{c}s_{\nu} \ , \cr
\frac{\partial F_{\nu\mu}}{\partial x_{\lambda}} + \frac{\partial F_{\mu\lambda}}{\partial x_{\nu}} +\frac{\partial F_{\lambda\nu}}{\partial x_{\mu}} &=& 0 \ .
\end{eqnarray} 
Here $s_{\nu} = (\mathbf{J}_{f}, ic\rho_{f})$ is the free-current four-vector.

To obtain (\ref{Forma4D}) one uses that 
\begin{eqnarray}
\frac{1}{c}g_{\nu} &=& \sum_{\mu = 1}^{4}\frac{\partial}{\partial x_{\mu}}M_{\nu \mu} \ ,
\end{eqnarray}
where $g_{\nu}$ is the bound-current four-vector given by
\begin{eqnarray}
g_{\nu} &\equiv& (\mathbf{J}_{b}, \ ic \mathbf{\rho}_{b}) \ , \cr
&=& \left( c\nabla \times\mathbf{M} + \frac{\partial \mathbf{P}}{\partial t}, \ -ic\nabla \cdot \mathbf{P} \right) \ .
\end{eqnarray}

Notice that $M_{\nu \mu}$ is obtained from $F_{\nu \mu}$ by replacing $B_{j}$ by $M_{j}$ and $E_{j}$ by $-P_{j}$, respectively. This difference in signs is responsible for the fact that the fields $\mathbf{E}$ and $\mathbf{B}$ transform with matrix $\mathsf{M}_{0}$, while the polarization $\mathbf{P}$ and magnetization $\mathbf{M}$ transform with the inverse matrix $\mathsf{M}_{0}^{-1}$, see equations (\ref{VC10}) and (\ref{MatrizPM}).


\end{document}